\documentclass[prd,aps,twocolumn,a4paper,showkeys,nofootinbib]{revtex4-1}
\usepackage{amsmath}
\usepackage{amsfonts}
\usepackage{amssymb}	
\usepackage{graphicx}
\usepackage{bm}
\usepackage{color}
\usepackage{commath}
\allowdisplaybreaks

\def\e{{\rm e}}

\def\GMc2{G M_{\odot} c^{-2}}

\def\lm{{\ell m}}

\def\lm{{\ell m}}

\def\lm{{\ell m}}

\def\ii{{\rm i}}

\def\F{{\cal F}}

\def\E{{\cal E}}

\newcommand\be{\begin{equation}}
\newcommand\ee{\end{equation}}

\usepackage[normalem]{ulem} 

\begin{document}
\title{Effective-one-body waveforms from dynamical captures in black hole binaries}
\author{Alessandro \surname{Nagar}${}^{1,2}$}
\author{Piero \surname{Rettegno}${}^{1,3}$}
\author{Rossella \surname{Gamba}${}^{4}$}
\author{Sebastiano \surname{Bernuzzi}${}^{4}$}
\affiliation{${}^1$ INFN Sezione di Torino, Via P. Giuria 1, 10125 Torino, Italy}
\affiliation{${}^2$ Institut des Hautes Etudes Scientifiques, 91440 Bures-sur-Yvette, France}
\affiliation{${}^{3}$ Dipartimento di Fisica, Universit\`a di Torino, via P. Giuria 1, 10125 Torino, Italy}
\affiliation{${}^{4}$Theoretisch-Physikalisches Institut, Friedrich-Schiller-Universit{\"a}t Jena, 07743, Jena, Germany}

\begin{abstract}
Dynamical capture is a possible formation channel for BBH mergers  
leading to highly eccentric merger dynamics and to gravitational wave
(GW) signals that are morphologically different from those of quasi-circular mergers.
The future detection of these mergers by ground-based or space-based 
GW interferometers can provide invaluable insights on astrophysical black
holes, but it requires precise predictions and dedicated waveform
models for the analysis. 
We present a state-of-the-art effective-one-body (EOB) model
for the multipolar merger-ringdown waveform from dynamical capture 
black-hole mergers with arbitrary mass-ratio and
nonprecessing spins. The model relies on analytical descriptions 
of the radiation reaction and waveform along generic orbits that 
are obtained by incorporating generic Newtonian 
prefactors in the expressions used in the quasi-circular case.
It provides a tool for generating waveforms for generic binary black 
hole coalescences and for GW data analysis.
We demonstrate that the model reliably accounts for the rich phenomenology
of dynamical captures, from direct plunge to successive close
encounters up to merger. The parameter space is fully characterized
in terms of the initial energy and angular momentum.
Our model reproduces to few percent the scattering angle from 
ten equal-mass, nonspinning, hyperbolic encounter numerical-relativity 
(NR) simulations. The agreement can be further improved to the 
by incorporating 6PN-results in one of the EOB potentials and
tuning currently unknown analytical parameters. Our results suggest 
that NR simulations  of hyperbolic encounters (and dynamical captures) 
can be used to inform EOB waveform models for generic BBH 
mergers/encounters for present and future GW detectors.
\end{abstract}
   
\date{\today}

\maketitle

\section{Introduction}
In dense stellar regions, e.g. galactic nuclei or globular clusters, individual black holes can 
become gravitationally bound as energy is lost to gravitational radiation during a close 
passage~\cite{Rasskazov:2019gjw, Tagawa:2019osr}.
Such dynamically captured pairs may be sources of gravitational radiation, with a 
phenomenology that is radically different from quasi-circular inspirals~\cite{Zevin:2018kzq,Samsing:2018isx}
and can be detected from larger distances and mass than 
quasi-circular mergers~\cite{10.1111/j.1365-2966.2009.14653.x}.
Although to date there is no observational evidence of these systems,
the capture of a stellar-mass object by a massive black hole
is also expected to be an efficient emitter of gravitational radiation for 
future detectors as the Einstein Telescope and LISA~\cite{Amaro-Seoane:2018gbb}.

Due to the special waveform morphology, these systems might be either missed
or incorrectly analyzed using standard quasi-circular templates, as already 
emphasized long ago~\cite{East:2012xq} (see Ref.~\cite{Loutrel:2020kmm} for
a recent review).  Physically faithful waveform 
models to systematically study the phenomenology of dynamical capture 
do not currently exists. To our knowledge, the only attempt at building a
waveform model for these kind of events dates back to Ref.~\cite{East:2012xq}
that provided a qualitative study of the phenomenon.
The model of~\cite{East:2012xq} is based on geodesic motion on a Kerr black hole 
spacetime augmented by leading-order Newtonian-like radiation reaction, then
complemented  by an effective model for a (quasi-circular) ringdown informed
by NR simulations\footnote{Note that the phenomenology of hyperbolic encounters/capture
i.e. {\it unbound} systems, is not accounted by currently available eccentric waveform
models like {\tt ENIGMA}~\cite{Huerta:2017kez} or {\tt SEOBNRE}~\cite{Cao:2017ndf,Liu:2019jpg}, 
that are limited to {\it bound} configurations with relatively mild eccentricity.}. 
Similarly, numerical relativity (NR) studies of BBH mergers from
dynamical capture conducted thus far are only
few~\cite{Gold:2012tk,Nelson:2019czq,Bae:2020hla} 
and limited to nonspinning binaries.

Reference~\cite{Damour:2014afa} shows that the effective one body (EOB) 
approach to the general relativistic two-body 
dynamics~\cite{Buonanno:1998gg,Buonanno:2000ef,Damour:2000we,Damour:2001tu,Damour:2008qf,Nagar:2011fx,Damour:2015isa} 
is suitable also for hyperbolic scattering events. In particular, Ref.~\cite{Damour:2014afa}
compares NR and EOB predictions for the scattering angles for
hyperbolic encounters, although it provides neither a description 
for the dynamical capture, nor a waveform model.
The aim of this paper is to go beyond the results of Ref.~\cite{Damour:2014afa} 
and illustrate that the EOB formalism can provide a complete model, for 
both dynamics and radiation, for dynamical capture black hole binaries. 
This model, which is publicly available as a stand-alone $C$ implementation~\cite{teobresums}, 
once optimized further, could be used for data analysis purposes, 
hopefully filling an evident gap in currently available waveform models. 
The key analytical advance used in this work 
is the radiation reaction and 
waveform along generic orbits proposed in
Ref.~\cite{Chiaramello:2020ehz}. The latter is based on the use of generic (noncircular) Newtonian prefactors in the (multipolar) 
waveform and radiation reaction. The merger and ringdown parts of 
the waveform are then modeled using analytical representations informed by 
quasi-circular (spin-aligned) numerical 
simulations~\cite{Nagar:2017jdw,Nagar:2019wds,Nagar:2020pcj}.
The approach developed here represents the state-of-art within the
currently available analytical knowledge.
However, it is currently impossible to make precise quantitative
statements on the actual faithfulness of the analytical waveforms
because of the lack of systematic predictions from NR simulations of 
dynamical capture coalescing binaries.
Moreover, the existing NR results, e.g.~\cite{East:2012xq,Gold:2012tk,Nelson:2019czq} 
are not expressed using gauge-invariant observables and so it is impossible to extract any kind of
information that could be directly compared with our model. Instead, we will provide
an up-to-date comparison with the measured scattering angle of Ref.~\cite{Damour:2014afa}.

The paper is organized as follows. In Sec.~\ref{sec:eob} we briefly review the eccentric  EOB 
model of Ref.~\cite{Chiaramello:2020ehz}, focusing in particular on the initial data setup,
as proposed already in Ref.~\cite{Damour:2014afa}. Section~\ref{sec:phenom} gives an overview
of the possible waveform phenomenology, considering both nonspinning and spinning cases
as well as higher multipolar modes. 
In Sec.~\ref{sec:chi} we build upon Ref.~\cite{Damour:2014afa}, providing a new comparison 
between EOB and NR scattering angles, with special emphasis on
the impact of recently computed 5PN and 6PN results~\cite{Bini:2019nra,Bini:2020wpo,Bini:2020nsb,Bini:2020hmy}.
In addition, differently from Ref.~\cite{Damour:2014afa} that was also relying on NR-computed 
fluxes, the computation of the scattering angle presented here is fully self-consistent within the
EOB formalism. Finally, in Sec.~\ref{sec:conclusions} we briefly summarize our findings and set 
the stage for future work. Geometric units with $G=c=1$ are employed unless otherwise specified.

\section{EOB waveform model for dynamical captures}
\label{sec:eob}
\subsection{Overview}
\label{sec:overview}
Reference~\cite{Chiaramello:2020ehz} introduced an EOB model valid along generic orbits,
i.e. valid for any configuration {\it beyond} the quasi-circular one. Although the analysis of Ref.~\cite{Chiaramello:2020ehz}
was limited to configurations with mild eccentricities ($\e\simeq 0.3$) there are no conceptual constraints
that prevent one from using the model in more extreme configurations, e.g. scattering and/or dynamical
captures. The eccentric EOB model we shall use here stems from the quasi-circular one
{\tt TEOBiResumS\_SM}~\cite{Nagar:2019wds,Nagar:2020pcj}, i.e. the most developed
version of {\tt TEOBResumS}~\cite{Nagar:2018zoe} that incorporates subdominant
waveform modes up to $\ell=m=5$. To generalize {\tt TEOBiResumS\_SM} to generic orbits,
Ref.~\cite{Chiaramello:2020ehz} proposed to simply replace the quasi-circular leading order terms, both 
in radiation reaction and waveform, with their  exact analytical expressions valid on general orbits.
The waveform is then completed using a model of (multipolar) merger and ringdown informed by
quasi-circular NR simulations~\cite{Damour:2014yha,Nagar:2020pcj}.
For all technical details we refer the reader to Refs.~\cite{Chiaramello:2020ehz,Nagar:2020pcj}.
We just recall the notation that will be useful in the reminder of the paper.
The two objects have masses $(m_1,m_2)$, with the convention that $m_1\geq m_2$ and the mass
ratio is defined as $q\equiv m_1/m_2\geq 1$. The total mass is $M=m_1+m_2$, the reduced mass
$\mu\equiv m_1 m_2/M$ and the symmetric mass ratio $\nu\equiv \mu/M$.
We use phase-space dimensionless variables $(r,p_{r_*},\varphi,p_\varphi)$, 
related to the physical ones $(R,P_{r_*},\varphi,P_{\varphi})$ by 
$r = R/GM$ (relative separation), $p_{r_*}= P_{R_*} / \mu$ (radial momentum),
$p_\varphi = P_\varphi / (\mu GM)$ (angular momentum), and $t \equiv T/(GM)$ the dimensionless time. 
The radial momentum $p_{r_*}$ is defined as $p_{r_*} = (A/B)^{1/2} \ p_r$, 
where $A$ and $B$ are the EOB metric potentials (see below).
The EOB (reduced) Hamiltonian describing the relative dynamics reads 
\be
\label{eq:H}
\hat{H}_{\rm EOB} \equiv H_{\rm EOB}\mu^{-1}= \nu^{-1}\sqrt{1+2\nu(\hat{H}_{\rm eff}-1)} ,
\ee
while the effective Hamiltonian $\hat{H}_{\rm eff}$ is split in a spin-orbit part (i.e. odd in spins)
and in the orbital part (i.e. even in spin)
\be
\label{eq:Heff}
\hat{H}_{\rm eff}=\hat{H}_{\rm SO}+\hat{H}_{\rm eff}^{\rm orb}.
\ee
The orbital part formally reads
\be
\label{eq:Heff_orb}
\hat{H}_{\rm eff}=\sqrt{A(1+p_\varphi^2 u_c^2 +Q)+p_{r_*}^2},
\ee
where $u_c\equiv 1/r_c$ and $r_c$ is the centrifugal radius that takes into account  
all effects even-in-spin mimicking the structure of the Hamiltonian of a particle on a
Kerr metric~\cite{Damour:2014sva}. Precisely following Ref.~\cite{Damour:2014sva},
let us recall that the $A$ potential is
\be
A = A_{\rm orb}(u_c;\nu)\dfrac{1+2 u_c}{1+2 u},
\ee
where $A_{\rm orb}$ is the orbital (nonspinning) potential that is used here at 5PNlog-accuracy,
resummed with a $(1,5)$ Pad\'e approximant, with an effective 5PN coefficient, $a_6^c$ informed
by NR simulations~\cite{Nagar:2020pcj}. Then, the other building elements, $(B,Q)$ are taken at 
(resummed) 3PN accuracy and are defined through
\be
D = A B = \dfrac{r^2}{r_c^2}D_{\rm orb}(u_c;\nu)
\ee
where
\be
D_{\rm orb}(u_c;\nu)=\dfrac{1}{1 + 6\nu u_c^2 + 2(26-3\nu)\nu u_c^3},
\ee
and $Q\equiv 2\nu(4-3\nu)p_{r_*}^4 u_c^2$. In the nonspinning case $r_c=r$, 
$D=D_{\rm orb}$ and $A=A_{\rm orb}$, and we will omit the subscript ``orb''
when discussing nonspinning configurations.

The waveform strain is decomposed in multipoles $h_\lm$ defined as
\be
\label{eq:hpc}
h_+ -\ii h_\times =\dfrac{1}{D_L}\sum_{\ell,m}h_{\ell m}\; {}_{-2}Y_{\ell m}(\theta,\Phi),
\ee
where ${D_L}$ is the distance from the source and  ${}_{-2}Y_{\ell m}(\theta,\Phi)$ 
are the $s=-2$ spin-weighted spherical harmonics. 
The model can generate all modes up to $\ell=m=5$ included, although, in the presence
of spin, the extension through merger and ringdown is not present for $(3,1)$, $(4,2)$ and $(4,1)$
(see discussion in Ref.~\cite{Chiaramello:2020ehz,Nagar:2020pcj}). 
Modes with $m=0$ are much smaller than the  $m\neq 0$ ones and so are currently 
omitted by the default EOB description\footnote{The $\ell=2$ $m=0$ mode can be approximately 
modeled within the current EOB framework as $h_{20}=-\dfrac{2}{7}\sqrt{\dfrac{10\pi}{3}}\nu(\dot{r} + r\ddot{r})\hat{H}_{\rm eff}$,
consistently with the 3PN order of the circularized case~\cite{Kidder:2007rt}. We have verified that for the
capture configurations considered in this paper the energy flux emitted during the dynamics due to this mode
is approximately two orders of magnitude smaller than for the $\ell=2$ mode, and thus negligible. The merger 
and ringdown part is currently not modeled. To do so, one will need future synergy with numerical results,
either in the extreme-mass-ratio limit, using BH perturbation theory~\cite{Harms:2014dqa}, 
or from NR simulations. }. 
\begin{figure*}[t]
\begin{center}
\includegraphics[width=0.32\textwidth]{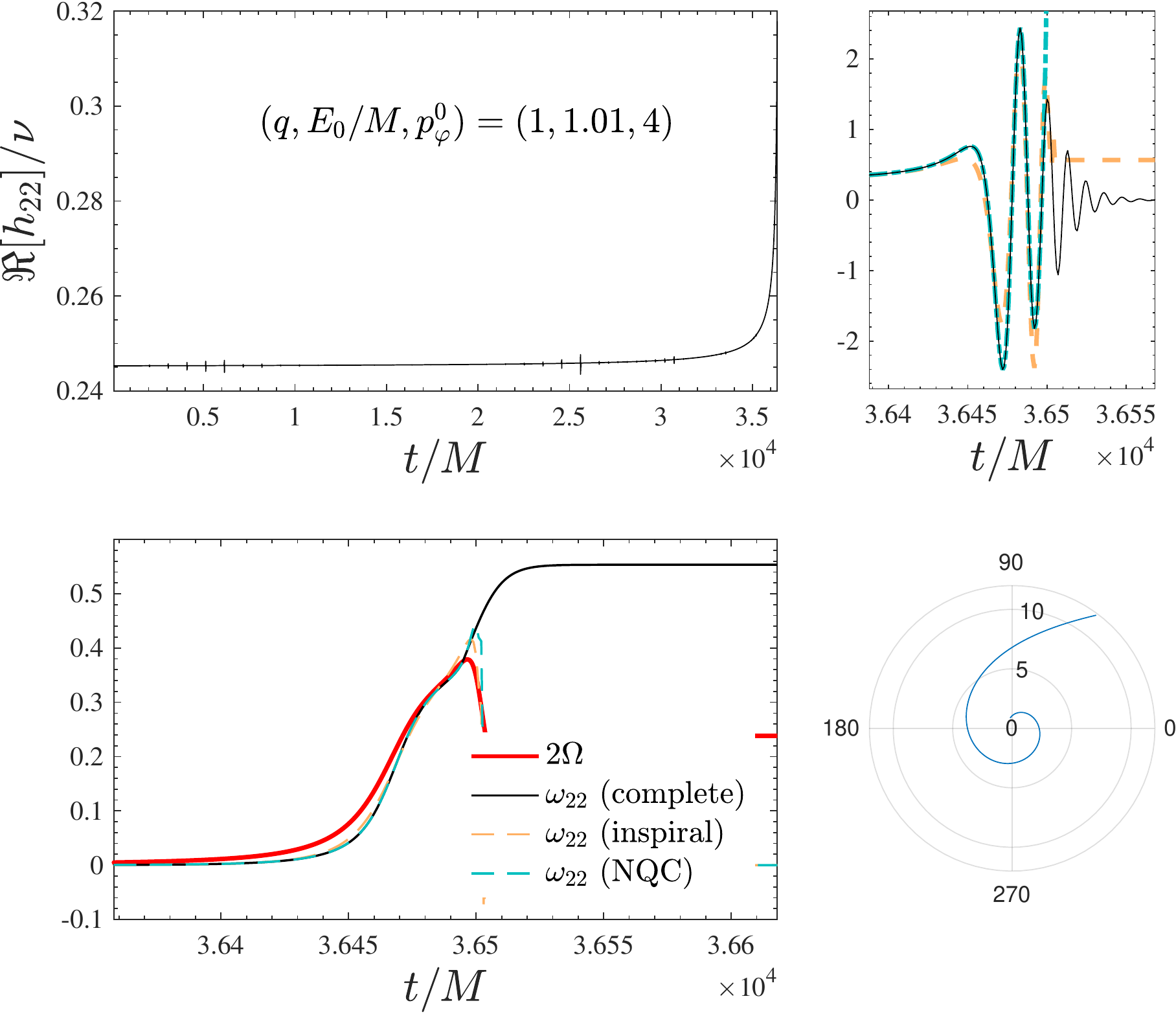}
\includegraphics[width=0.32\textwidth]{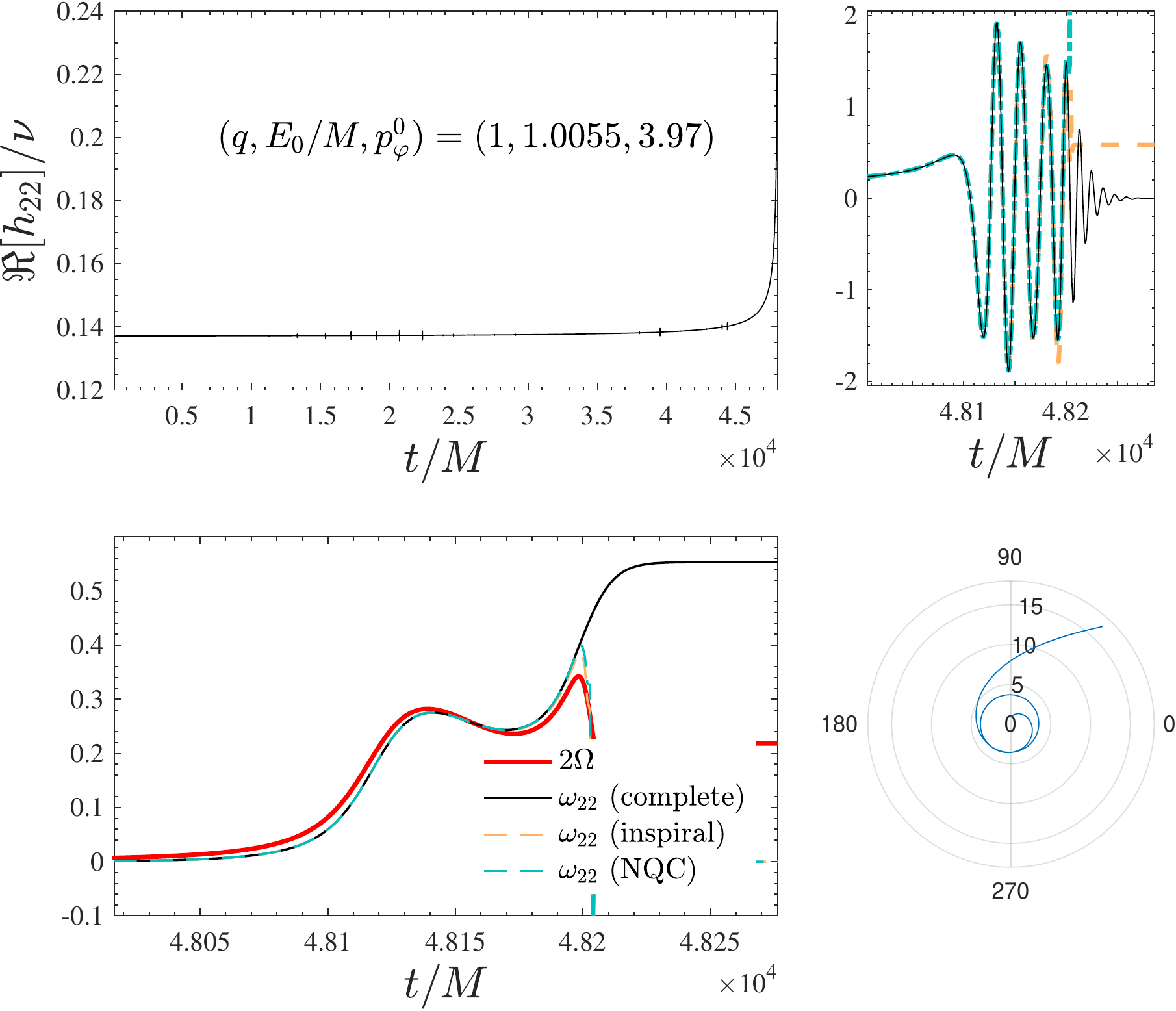}
\includegraphics[width=0.32\textwidth]{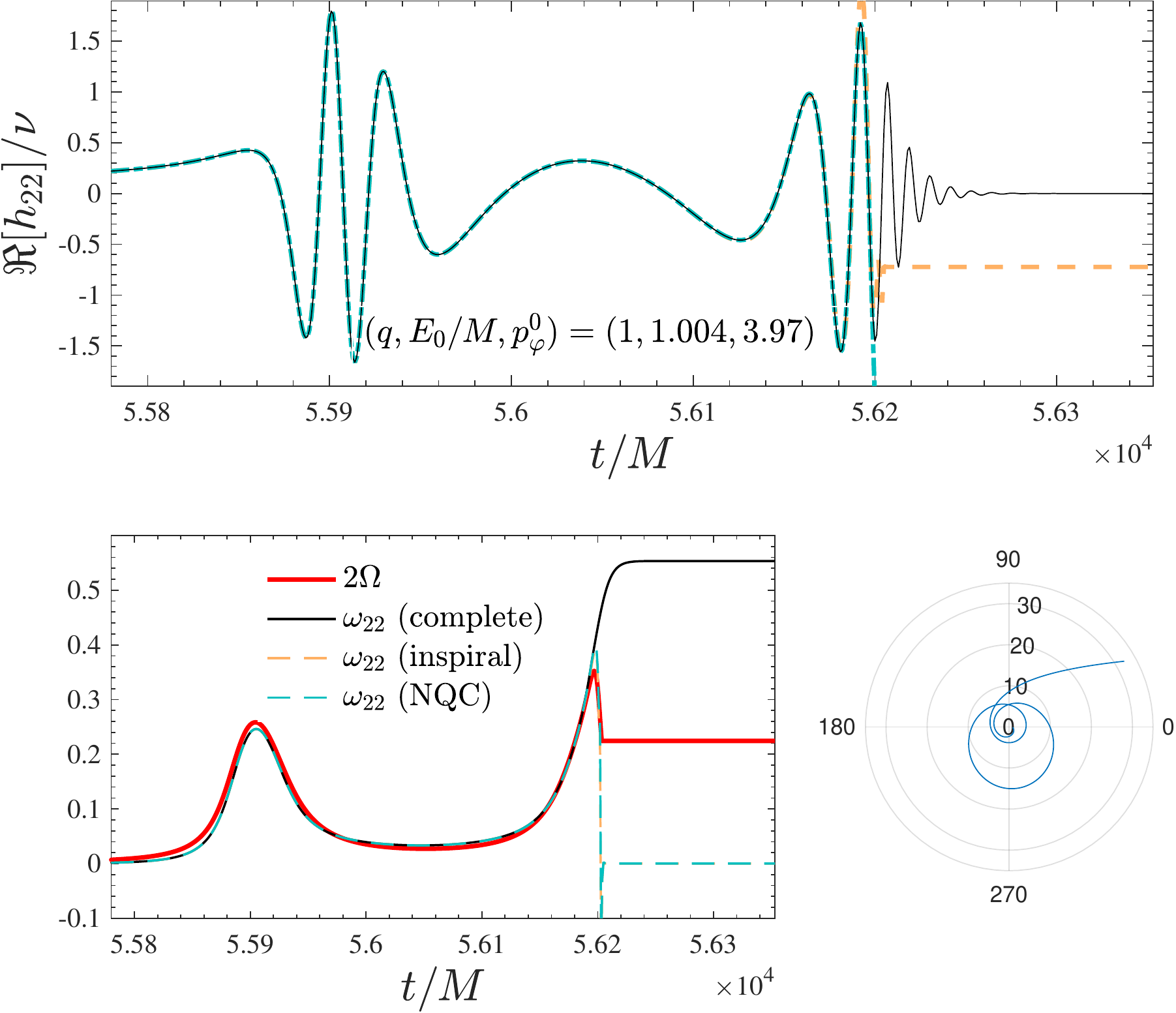}
\caption{Waveform phenomenology for $q=1$: selected configuration to illustrate the transition
from direct plunge (left panel) to one distinct encounter before the final merger (right panel).
Each configuration is characterized by initial data $(q,E_0/M,p_\varphi^0)$.
The top row of each panel shows the real part of the $\ell=m=2$ waveform, completed with merger and ringdown (black online), 
with close ups onto the final merger part. The left-bottom panel exhibits the gravitational wave 
frequency, while the right-bottom the last part of the relative separation $r(\varphi)$. 
The panels also show: (i) the purely analytical EOB (inspiral) waveform and frequency (orange online); 
(ii) the waveform completed by NR-informed next-to-quasi-circular corrections (NQC, blue online). 
Note that the analytical EOB waveform accounts for the GW emission up to the largest peak of 
the orbital frequency $\Omega$ (red online). It is only after that this point is reached that the
postmerger-ringdown description is attached, analogously to the quasi-circular case~\cite{Damour:2014sva,Nagar:2017jdw,Nagar:2018zoe,Nagar:2019wds,Nagar:2020pcj}.}
\label{fig:q1}
\end{center}
\end{figure*}

\subsection{Setup of initial data}
\label{sec:ID}
The initial data setup for hyperbolic encounters/dynamical capture was already discussed
in Ref.~\cite{Damour:2014afa}. One proceeds as follows: (i) fix a value of the angular momentum
$p_\varphi$ such that the local peak of the potential energy
is {\it larger} than $1$ (see Fig.~3 of Ref.~\cite{Damour:2014afa} or
Fig.~\ref{fig:q1spin} below); (ii) choose a value of the initial energy
$E_0/M$; (iii) choose a value of the initial separation, that should be
large, typically $r_0=10000$ (though sometimes we also use $r_0=1500$); 
(iv) solve Eq.~\eqref{eq:H} for $p_{r_*}$  with $H_{\rm EOB}=E_0$. Since at such
large values of $r_0$ spin effects are negligible, we can simply consider
the nonspinning version of Eqs.~\eqref{eq:Heff}-\eqref{eq:Heff_orb},
with $r_c=r$~\cite{Damour:2014sva} to obtain $p_{r_*}$. To span the parameter
space, we first fix $p_\varphi=p_\varphi^0$ and we then choose the initial energy
between $E_{\rm max}/M$, the energy of the unstable circular orbit
(i.e. the peak of the effective potential $A(1+p_\varphi^2 u^2)$)
and $E_{\rm min}/M = H_{\rm EOB}(p_\varphi^0,p_{r_*}=0,r_0)/M$. As discussed
below, the region of parameter space with $E_0\leq E_{\rm max}$ is the one
with the most interesting phenomenology, i.e. with the possibility of
having many close passages before merger. On the contrary, the regions
with $E_0\geq E_{\rm max}$ always correspond to direct capture events.
We will not discuss explicitly this phenomenology, though it is obviously
also allowed by the formalism.

\section{Dynamical capture phenomenology}
\label{sec:phenom}
Let us now give a general overview of the properties of the relative dynamics and
waveforms from dynamical capture as predicted by our EOB model. Note that we
will mainly focus on the capture scenario and  discuss in Sec.~\ref{sec:chi} below 
the scattering scenario.
To simplify the discussion, we start by considering the $q=1$, nonspinning case.
To setup initial data, we consider values of the angular momentum $p_\varphi$ 
sufficiently larger than the LSO value, $p_\varphi^{LSO}$ so as to allow for the 
peak of the potential energy to be larger
than one\footnote{For the nonspinning case, from the
conservative EOB Hamiltonian one obtains $p_\varphi^{\rm LSO}(\nu)=3.4643-0.774482\nu-0.692\nu^2$.}.
For each value of $p_\varphi$ we select values of the energy between
$(\hat{E}_{\rm min},\hat{E}_{\rm max})$ as mentioned above.
At a qualitative level, for a given value of $p_\varphi^0$, as the energy is decreased 
from $\hat{E}_{\rm max}$, the system passes through the following stages: 
(i) direct capture/plunge; (ii) one, or more, close encounters before merger; 
(iii) close passage and scattering away.
In practice, the detailed behavior as energy is decreased is more complicated,
because, as $E\to 1$ the system moves from scattering configurations back to
(many) close encounters that eventually end up with gravitational capture.
More details on this phenomenology will be given below.
\begin{figure*}[t]
  \begin{center}
  \includegraphics[width=0.22\textwidth]{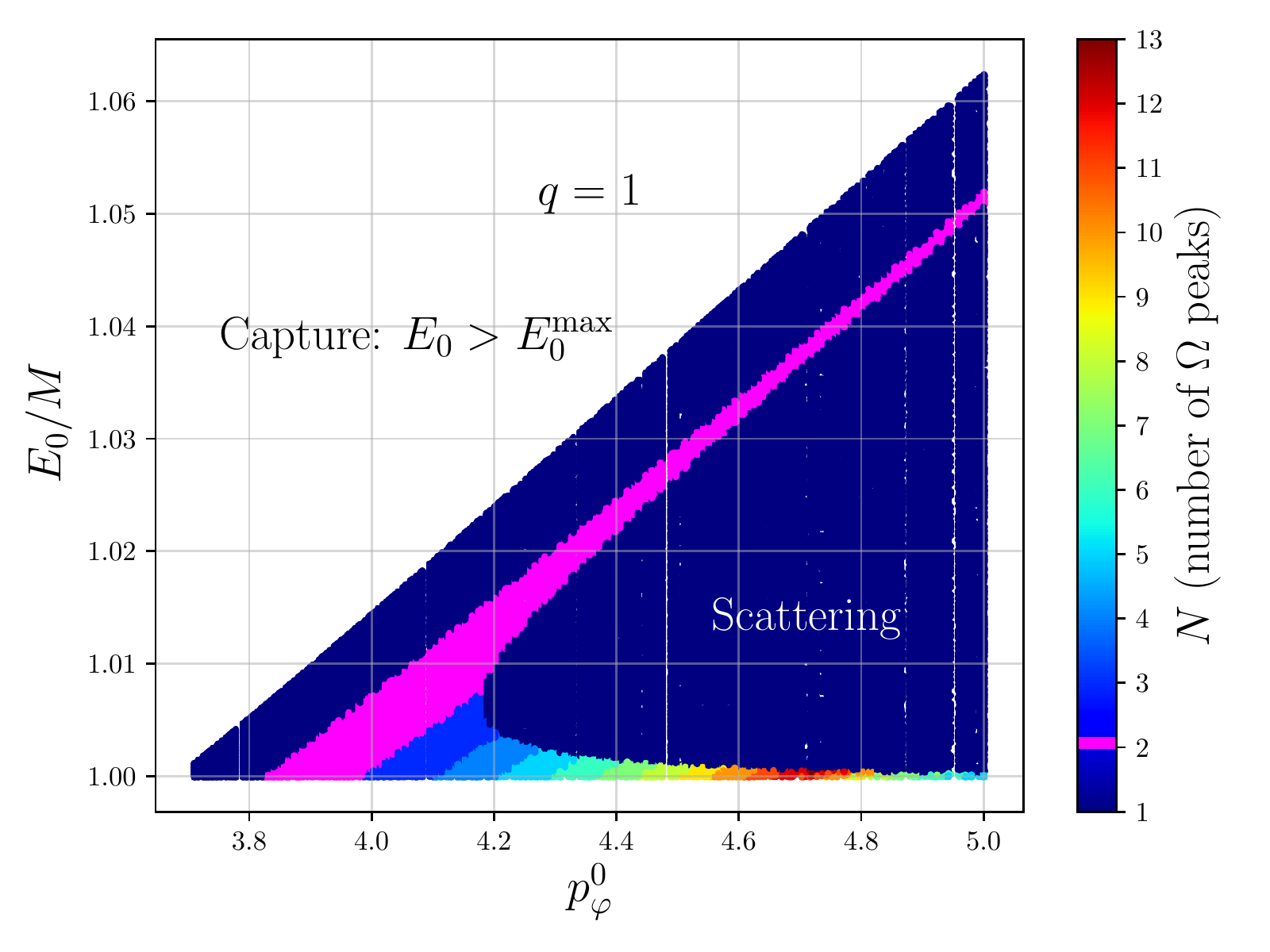}
  \includegraphics[width=0.22\textwidth]{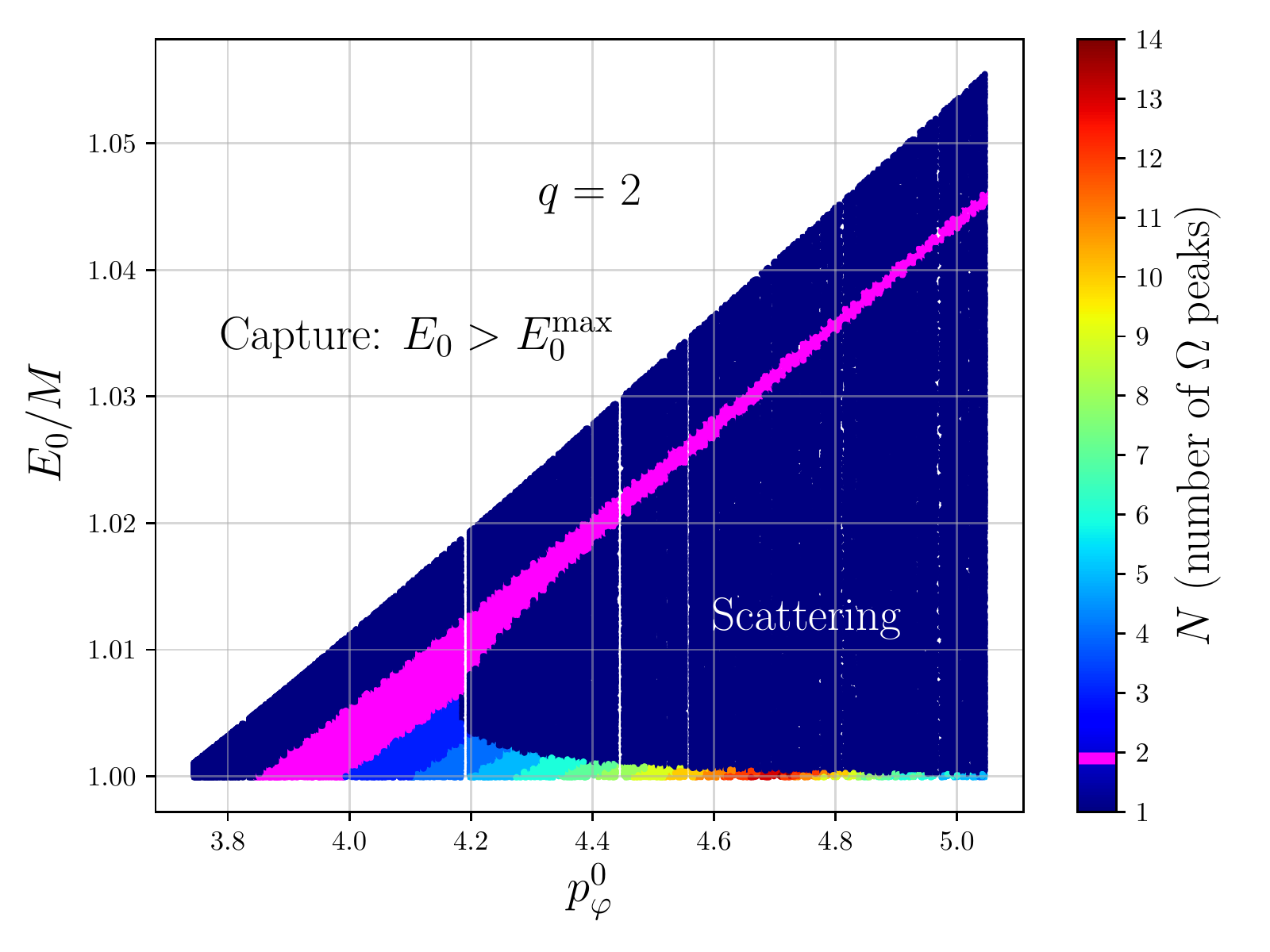}
  \includegraphics[width=0.22\textwidth]{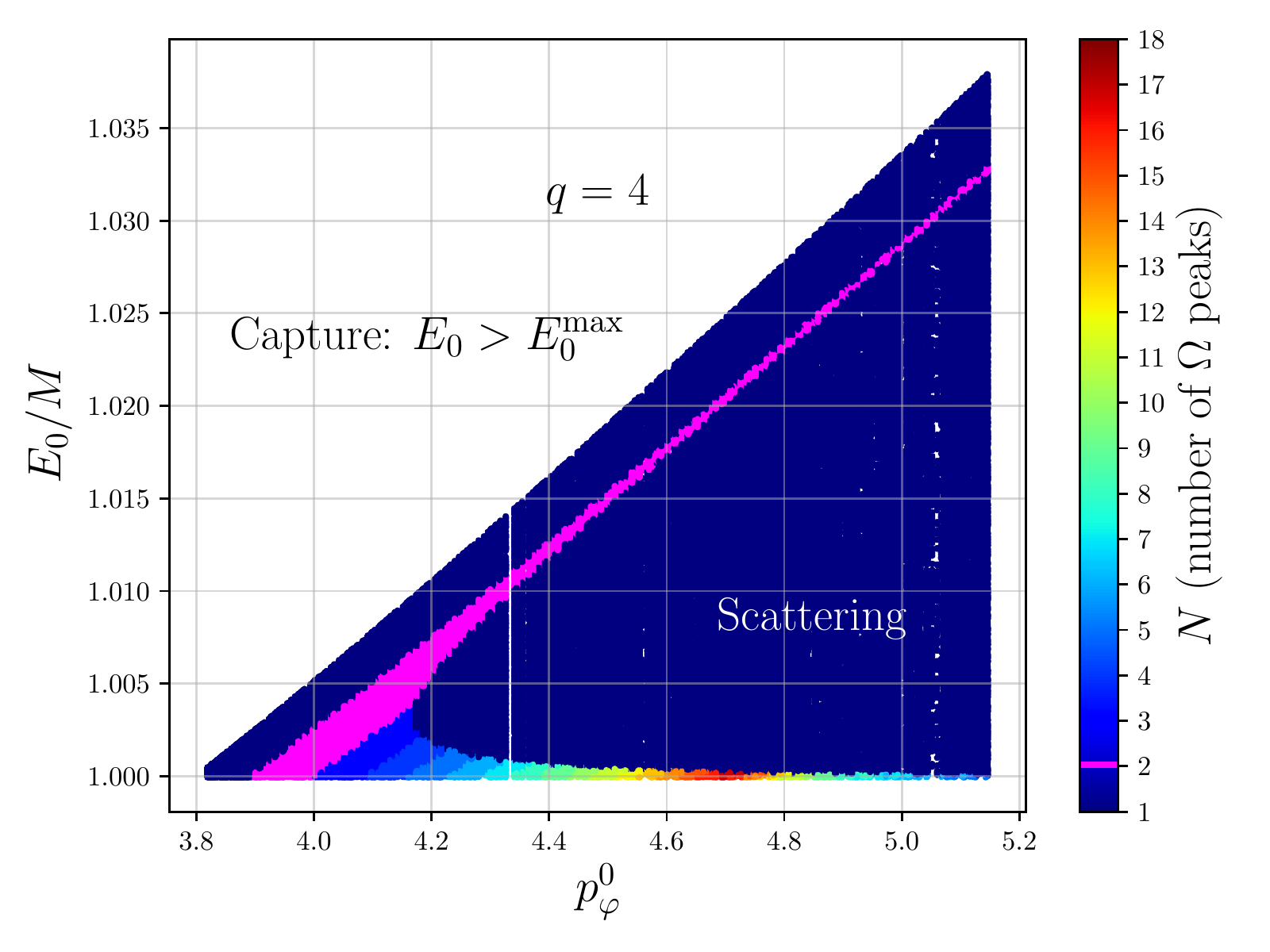} 
  \includegraphics[width=0.22\textwidth]{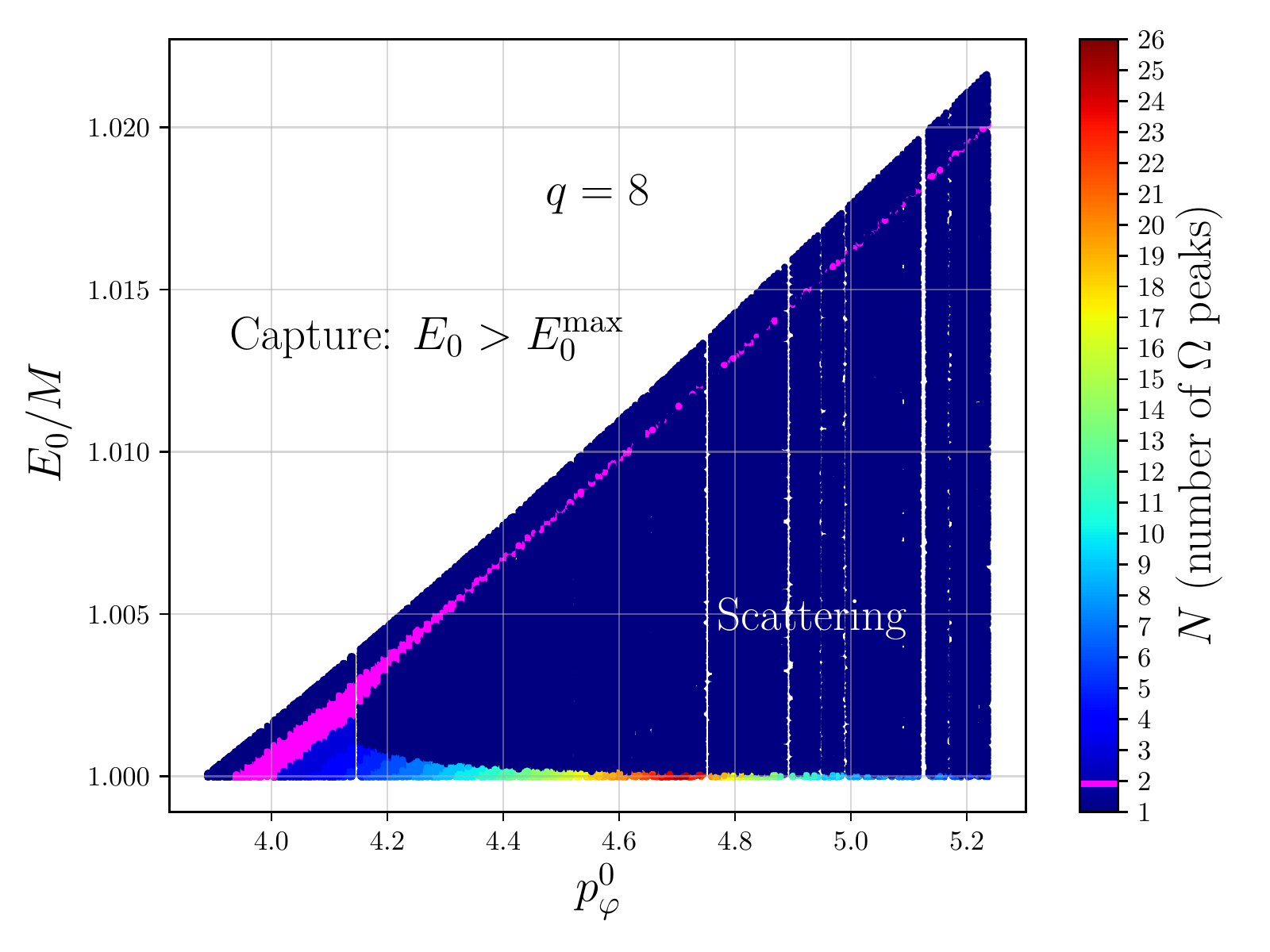}\\
  \includegraphics[width=0.22\textwidth]{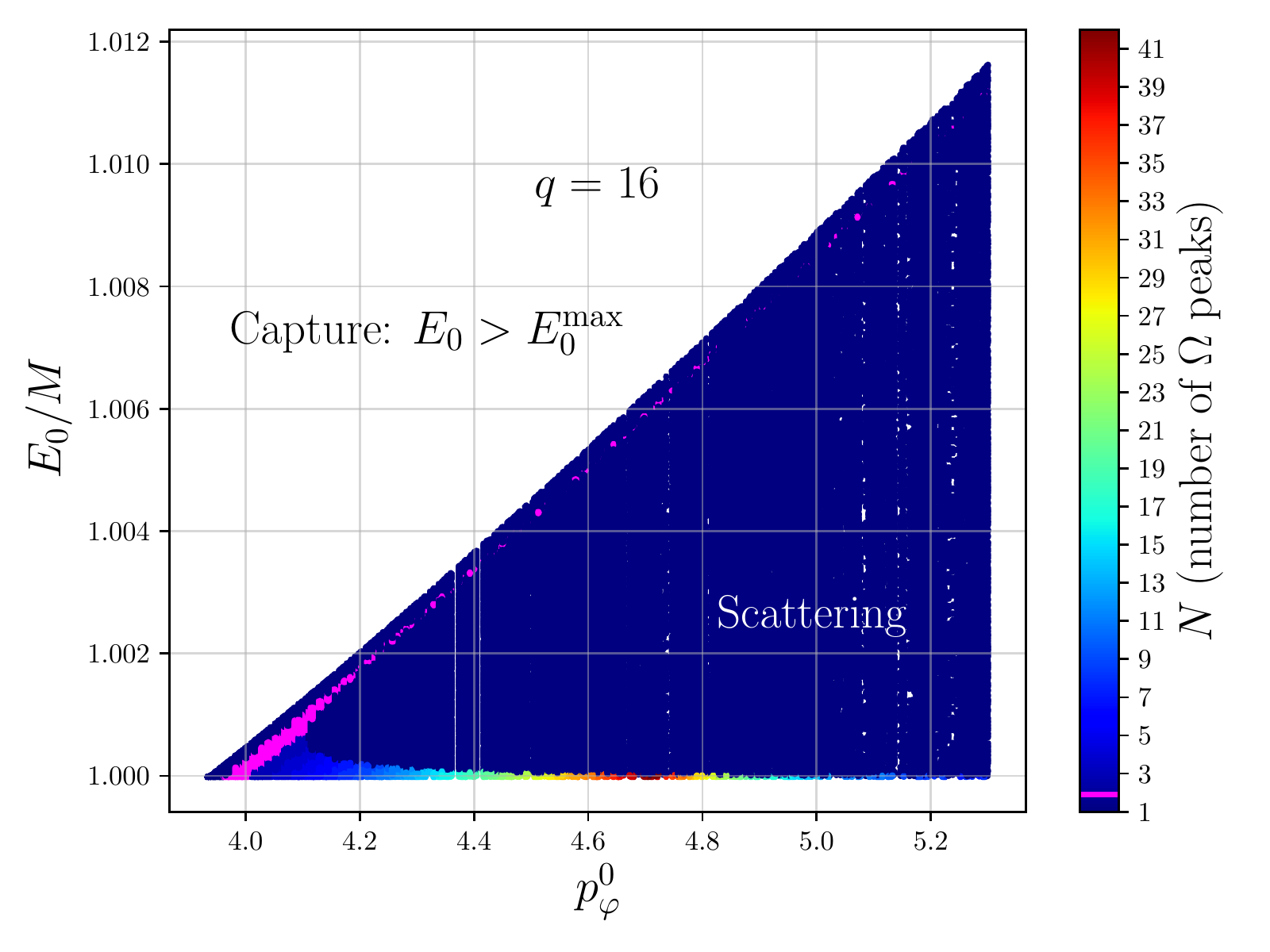}
  \includegraphics[width=0.22\textwidth]{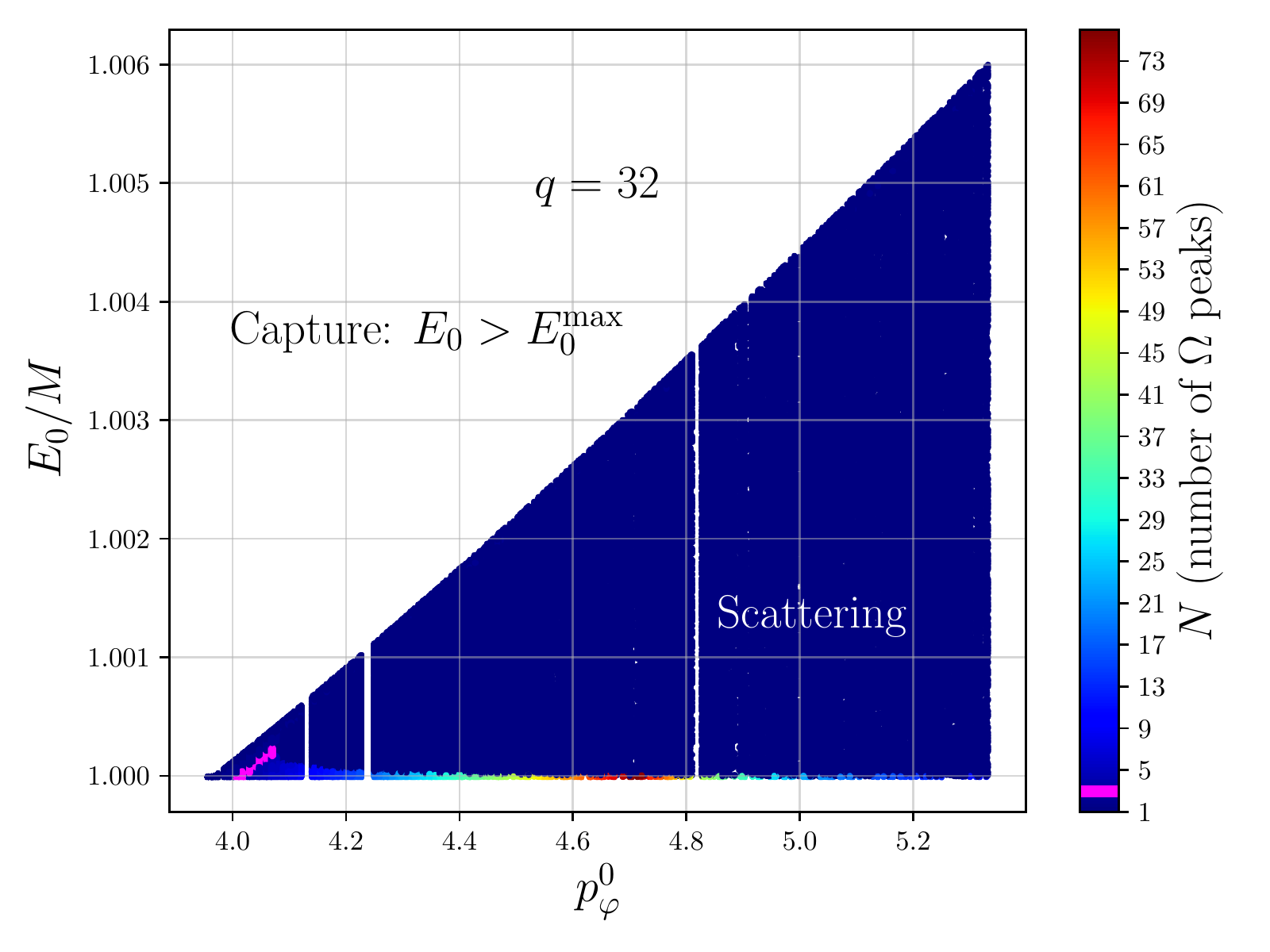}
  \includegraphics[width=0.22\textwidth]{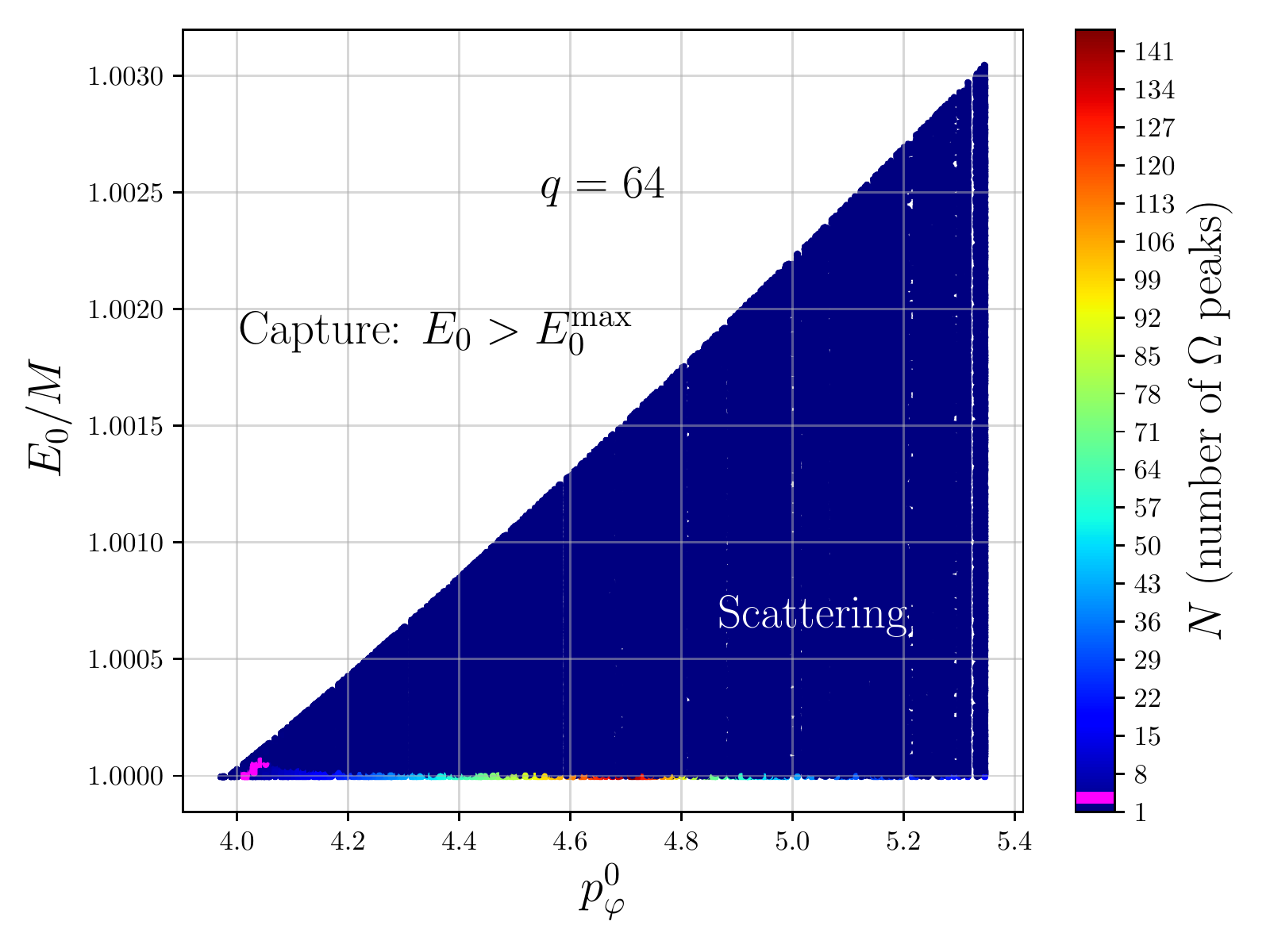}
  \includegraphics[width=0.22\textwidth]{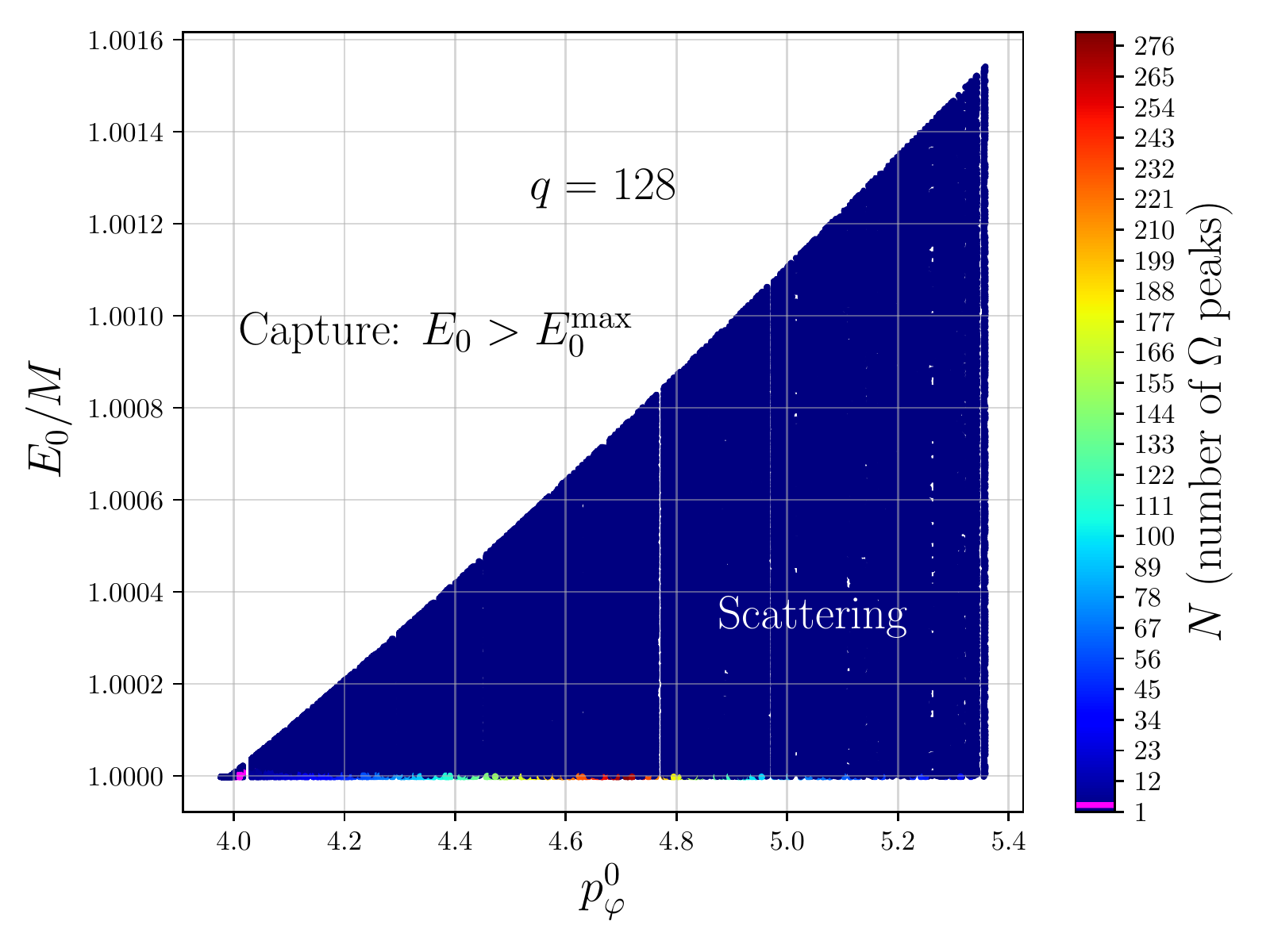}    
\caption{Analysis of the parameter space of hyperbolic encounters of nonspinning BBHs parameterized in terms 
of initial data $(q,E_0/M,p_\varphi^0)$. The number of multiple encounters ($N\geq 2$) increases with 
$q$, while the corresponding area on the parameter space gets smaller and smaller. Note the separation, 
given by the colored area, between configurations that scatter and configurations that eventually merge.}
\label{fig:survey}
\end{center}
\end{figure*}
\begin{figure}[t]
\begin{center}
\includegraphics[width=0.45\textwidth]{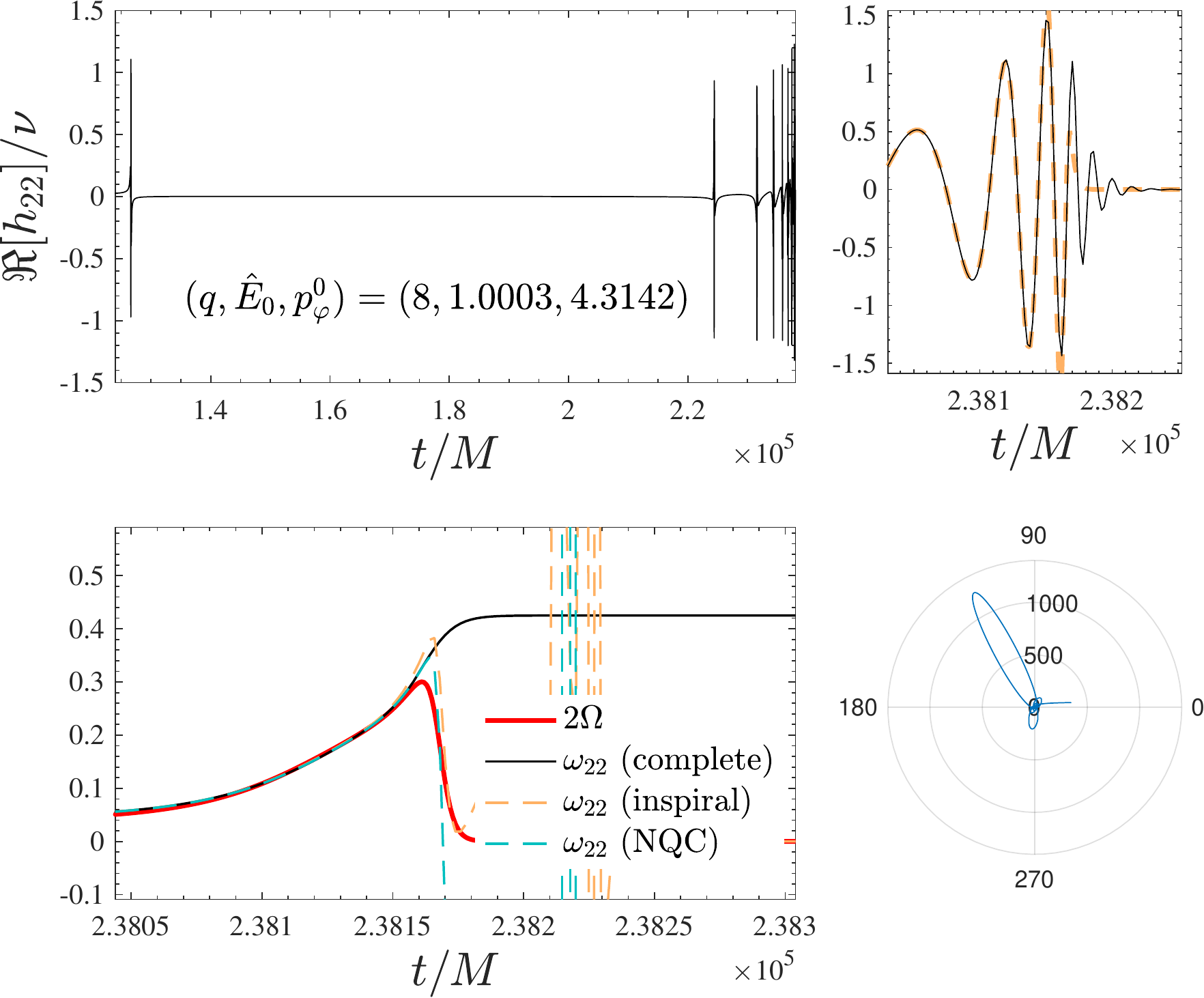}
\caption{Extreme configuration with $(q,E_0/M,p_\varphi^0)=(8,1.0003,4.3142)$ that 
undergoes eight periastron passages before merger.   Note, after the first encounter, 
the very eccentric orbit, with apastron that reaches $r\simeq 1200$ (bottom left panel).
Top right panel: close up on the final plunge, merger and ringdown part of the waveform. 
For completeness we also incorporate the purely analytical waveform (orange).}
\label{fig:q8config}
\end{center}
\end{figure}

To start with, Fig.~\ref{fig:q1} shows three waveforms with nearly the same value of the angular
momentum, but where the energy is progressively decreased. The configurations were selected
so that one can appreciate the transition from immediate scattering (left panel) to a quasi-circular
capture (middle panel), where the system does a full quasi-circular orbit before plunging, and
the case when there's a close encounter followed by capture and merger (right panel).
For each configuration, we show: (i) the real part of the waveform; (ii) the gravitational wave
frequency $\omega_{22}$ together with twice the orbital frequency $2\Omega$; and (iii) the
orbit $r(\varphi)$ of the relative separation.  
For completeness, in  both the waveform and frequency panels we include three curves:
(i) the simple, analytical,  EOB waveform, with the general Newtonian prefactor
as explained in Ref.~\cite{Chiaramello:2020ehz} (dashed, orange); (ii) the waveform corrected by additional
next-to-quasi-circular (NQC) factors, that are informed by quasi-circular NR simulations following
now standard procedures (light blue, dash-dotted) and the waveform completed with the, 
similarly NR-informed, ringdown. More precisely, the ringdown is attached at $t=2M$ after
the peak of the $\ell=m=2$ analytic waveform, according to the standard procedure implemented
in the various flavors of {\tt TEOBResumS}~\cite{Nagar:2020pcj,Chiaramello:2020ehz}.
To characterize the dynamics, it is useful to look at the morphology of the orbital frequency.
In the case of immediate plunge, $\Omega=\dot{\varphi}$ has a single peak, corresponding to the crossing
of the EOB effective light-ring. When the energy is lowered (see middle panel of Fig.~\ref{fig:q1}),
the frequency progressively flattens and an {\it earlier peak} appears well before the merger one.
This ``precursor'' peak corresponds to a periastron passage with $r_{\rm periastron}\simeq 3.47$;
after this, $r$ increases again and eventually the system plunges, with a second peak in $\Omega$.
As the energy is further lowered (see third panel of Fig.~\ref{fig:q1}), the first peak, that
corresponds to the first close passage, becomes clearly distinguishable and separate from
the one corresponding to merger.  Inspecting the middle panel of Fig.~\ref{fig:q1}, one then
understands that the divide between having an immediate plunge and a close encounter followed
by a plunge is determined by the condition $\dot{\Omega}=\ddot{\Omega}=0$, i.e. the orbital
frequency should have an inflection point at some time (radius).
Inspecting the left panel of Fig.~\ref{fig:q1}, one sees how the late, inspiral-like, part
of the orbit is fully mirrored by one entire GW cycles in the purely analytical waveform
{\it before} the ringdown signal actually occurs. Similarly, in the middle panel one sees
that the waveform mirrors the quasi-circular dynamics giving four, entire, GW cycles before
merger and ringdown. 

On the basis of the morphological analysis of above, a simple way to characterize the
parameter space of dynamical capture is to focus on the orbital  frequency
as function of time, $\Omega(t)$, and count how many peaks are present.
A single peak may correspond to either immediate plunge or scattering.
More generally, when many peaks are present, each peak corresponds
to a periastron passage. So, the number of peaks of $\Omega(t)$ is a
simple observable, function of $(q,E_0/M,p_\varphi^0)$ that could
be used to characterize the parameter space of dynamical capture BBHs~\footnote{An 
equivalent  observable is given by the number of peaks of the gravitational wave frequency,
any isolated peak corresponding to a periastron passage. Using the GW frequency has
the advantage that the analysis we are discussing here can be directly extended to 
NR simulations, using then the same peak number as function of initial ADM energy
and angular momentum to fully characterize the parameter space.}.

We then consider different mass ratios, $q=\{1,2,4,8,16,32,64,128\}$ to provide 
a comprehensive mapping of the parameter space. For each value of the angular 
momentum, we lower the energy and count the number of peaks of $\Omega$. 
The result of this analysis is reported in  Fig.~\ref{fig:survey}.
The colors characterize how many periastron passages the system has
undergone before merging.
Focusing first on the $q=1$ case (top-left panel of the figure), one sees that 
when the energy is decreased from $\hat{E}_{\rm max}$ there are different 
islands of initial parameters that correspond to progressively more complicated 
physical behaviors. The plot is split in two by an area that corresponds to 
the frequency developing two peaks before merger (magenta online).
As mentioned above, The upper boundary of this region is defined by those
values  of $(E_0/M,p_\varphi^0)$ such that $\dot{\Omega}=\ddot{\Omega}=0$
at some time. Now, the $N=1$ part of the parameter space {\it above} the magenta region
corresponds to direct  plunge, with a waveform phenomenology similar to the one in
the left panel of Fig.~\ref{fig:q1}. By contrast, the $N=1$  part on the right and below
the magenta region corresponds to scattering events instead of capture. 
When the initial energy is lowered further, getting close to the stability region, 
the system attempts to stabilize again and the number of periastron passages  before 
merger increases progressively also for large values of $p_\varphi^0$. 
The phenomenology remains qualitatively the same {\it also} 
when the mass ratio is increased, but the region with $N=2$ becomes narrower and narrower
as $q$ increases, notably for $q\geq 32$, when the divide between $N=1$ configuration is barely
visible on the plots (we shall quantify this behavior better below).
By contrast, for energies just slightly larger than the (adiabatic) stability limit, the number 
of possible encounters can grow considerably, up to several tens, although limited to a 
region of $p_\varphi^0$ much smaller than in the equal-mass case. We qualitatively interpret 
this behavior as mirroring the effect that radiation reaction, that is proportional to $\nu$,
becomes less and less efficient as $\nu$ is decreased and so the system can 
persist in  a metastable state much longer.
\begin{figure}[t]
\begin{center}
\includegraphics[width=0.45\textwidth]{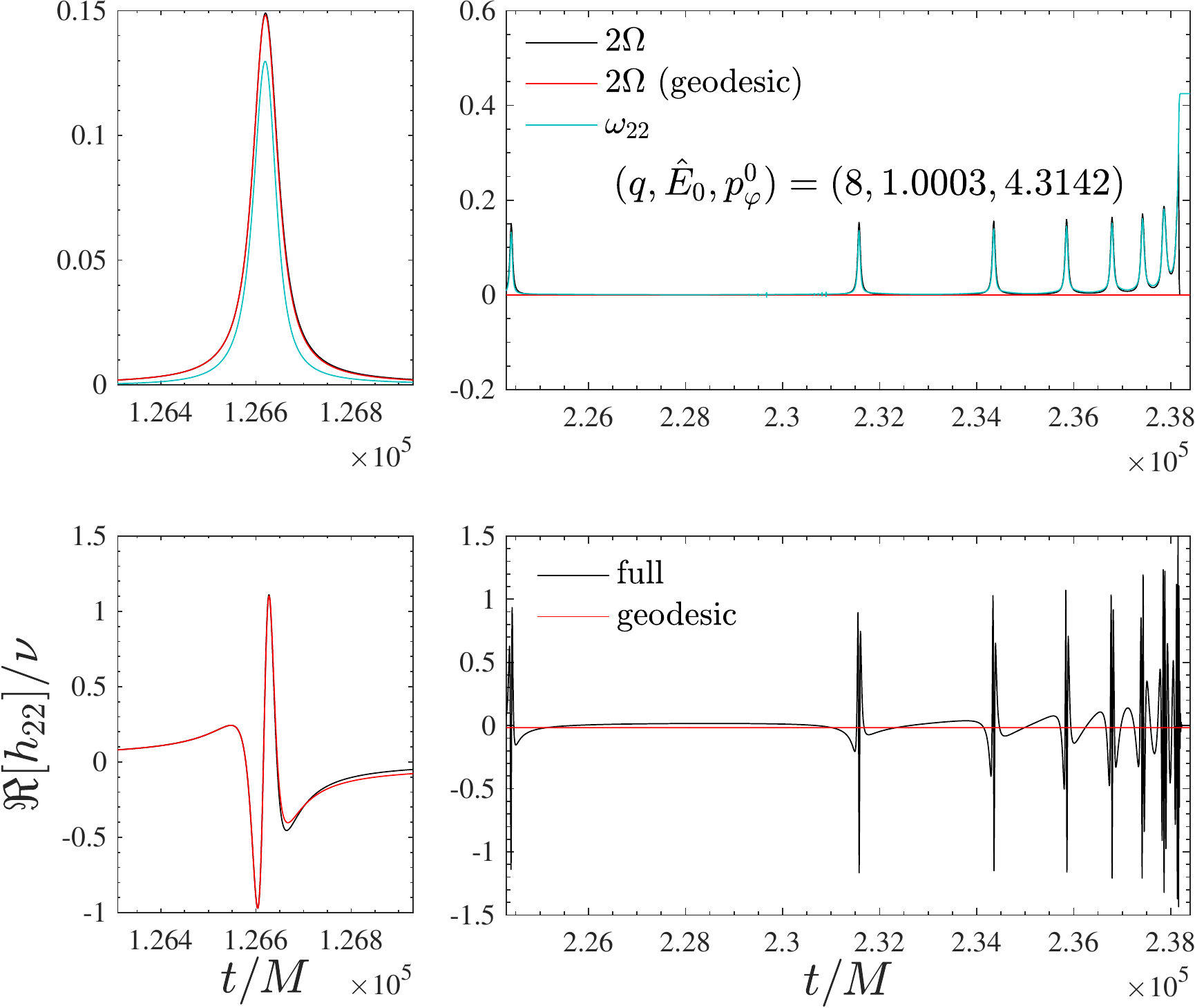}
\caption{Importance of radiation reaction to determine the multiple-encounters behavior of Fig.~\ref{fig:q8config},
with the same initial conditions $(q,E_0/M,p_\varphi^0)=(8,1.0003,4.3142)$ considered there. The black lines
correspond to the full dynamics with radiation reaction, while the red ones correspond to the  conservative 
dynamics alone.}
\label{fig:q8_geo}
\end{center}
\end{figure}

In order to give an explicit example of the complicated phenomenology of a capture that occurs after
many close encounters, let us consider a configuration with $(q,E_0/M,p_\varphi^0)=(8, 1.00026983016,4.3141870095)$,
that is exhibited in Fig.~\ref{fig:q8config}. The top row of the figure shows the real part of the 
waveform, each burst corresponding to a close passage. Analogously to what done in Fig.~\ref{fig:q1},
the close up of the waveform  around merger (top-right panel) also includes the analytical
EOB waveform (orange). The bottom row exhibits the time-evolution of the frequency around merger as well as
the relative trajectory. It is remarkable that after the first encounter the system 
undergoes an extremely elliptic orbit, with apastron reaching $r\simeq 1200$,
before being captured again. This behavior is determined by the action of radiation reaction
around the first close encounter: in that situation the system emits a burst of radiation that
eventually makes the orbit close again instead of scattering away. We prove
this by setting up the EOB dynamics with the same initial data, but switching off radiation
reaction, i.e. both $\hat{\cal F}_\varphi=\hat{\cal F}_{r_*}=0$. Figure~\ref{fig:q8_geo} compares
the orbital frequency and waveform of Fig.~\ref{fig:q8config} with the waveform obtained from
the conservative dynamics only. The location of the first burst, that  corresponds to the first
encounter, is essentially the same for both configurations, highlighting that the effect of
radiation reaction is practically negligible up to that point. The differences in the
waveform (and thus dynamics) occur later, consistently with the fact mentioned above that the
effect of radiation reaction is localized around the periastron passage. In the top panel of the figure 
we also compare the full GW frequency, $\omega_{22}$ with twice the orbital frequency $\Omega$,
to highlight that $\omega_{22}\neq 2\Omega$ because of the various noncircular effects occurring
near the periastron. From this example one also argues that the span of the capture region in the
parameter space depends on the details of the model for radiation reaction and may thus 
change if an improved version of the latter is implemented within the model~\footnote{In this respect,
let us recall that the time-derivatives of $(r,\Omega)$ entering the generic Newtonian prefactor in the radiation
reaction are obtained following a certain iterative procedure discussed in~\cite{Chiaramello:2020ehz}
and based upon results of Appendix~A of Ref.~\cite{Damour:2012ky}. Consistently with 
Ref.~\cite{Chiaramello:2020ehz}, we work here with 2 iterations, that are sufficiently accurate 
for mild eccentricities. We have however verified that the configuration of Fig.~\ref{fig:q8config}, 
because of the highly-eccentric orbit following the first encounter, is sensitive to these details and 
one obtains quantitatively different waveforms when using $2$ or $4$ iterations: the waveform in 
the second case is a bit longer. Despite this, the number $N$ of peaks of $\Omega$, as well as 
the phenomenology discussed so far, do not change.}. We will comment more on the 
issue of {\it analytical uncertainty} of the model,
and the related importance of NR simulations as a benchmark, in Sec.~\ref{sec:chi} below. 

\begin{figure}[t]
\begin{center}
\includegraphics[width=0.45\textwidth]{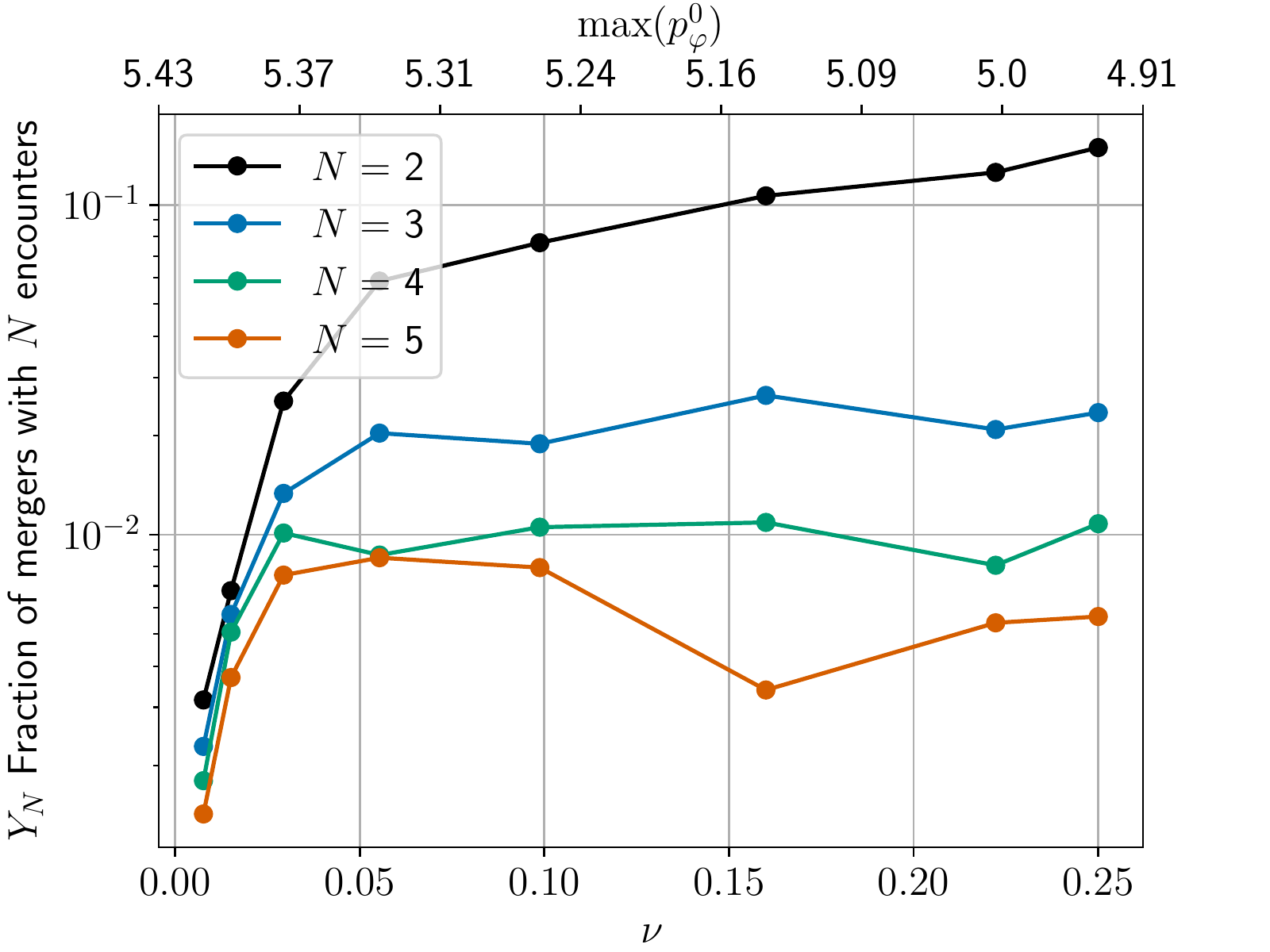}
\caption{Fraction of BBH configurations (including also scattering events) that end up with $N$ encounters 
(where the $N$-th encounter corresponds to merger) for nonspinning binaries. Configurations with 
$N=2$ are the most frequent ones, although their frequency quickly decreases 
below $10\%$ as $q>4$ ($\nu <0.16$).}
\label{fig:nospin_fraction}
\end{center}
\end{figure}

Finally, to put on a more quantitative ground the specific qualitative observations made so far,
we compute, for each value $q=\{1,2,4,8,16,32,64,128\}$ the fraction $Y_N$ of events with 
$N$ encounters (where the $N$-th encounter corresponds to merger in case of final capture).
Figure~\ref{fig:nospin_fraction} exhibits this quantity versus $\nu=q/(1+q)^2$. Configurations 
with two encounters are always the most frequent ones, although their fraction quickly 
decreases below $10\%$ for $q>4$ ($\nu <0.16$).

\subsection{Spin}
\label{sec:spin}
\begin{figure}[t]
\begin{center}
\includegraphics[width=0.45\textwidth]{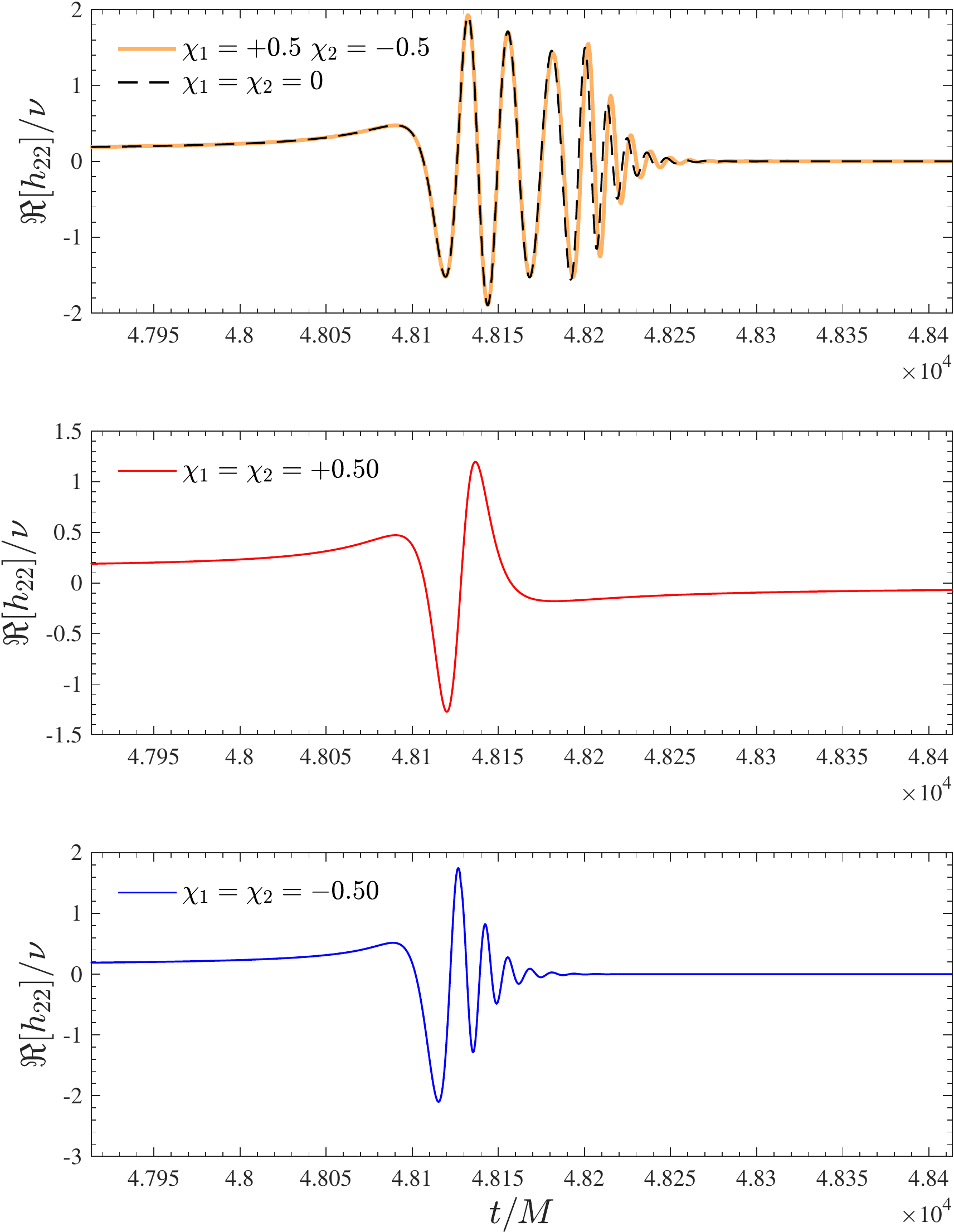}
\caption{Effect of the spins on the $(q,E_0/M,p_\varphi^0)=(1,1.0055,3.97)$ configuration discussed above. 
Top panel: when $\chi_1$ and $\chi_2$ are anti-aligned among themselves, the spin-orbit interaction cancels out and the
waveform is almost equivalent to the nonspinning one. Middle panel: the repulsive character of spin-orbit interaction
for spins aligned with the angular momentum is such to have a scattering instead of a dynamical capture.
Bottom panel: when spins are anti-aligned with the angular momentum, the system plunges faster, with 
a short burst of radiation corresponding to the final capture.}
\label{fig:q1spin}
\end{center}
\end{figure}

\begin{figure}[t]
\begin{center}
\includegraphics[width=0.45\textwidth]{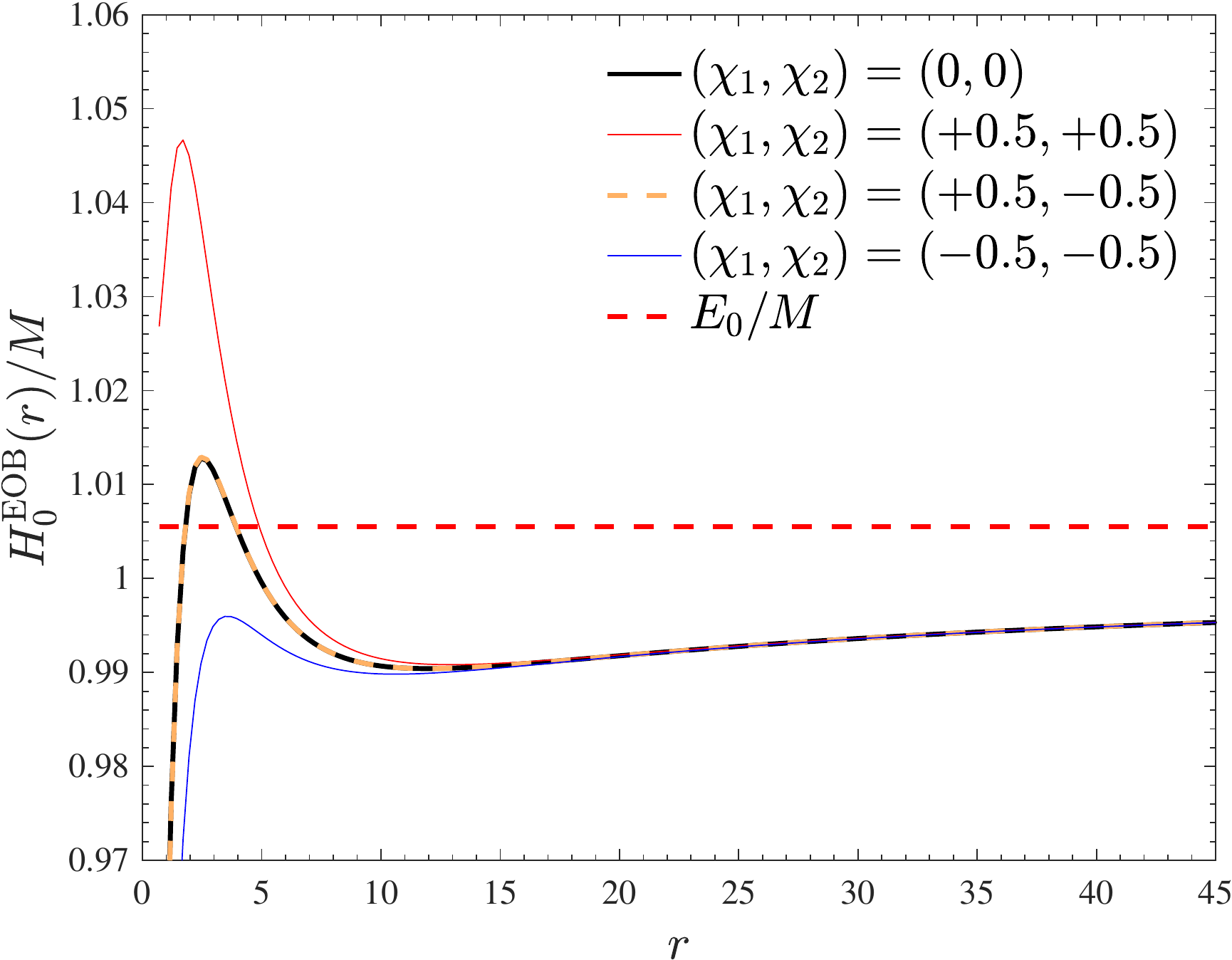}
\caption{Potential energy $H^{\rm EOB}_0(r)/M$ for the four configurations shown in Fig.~\ref{fig:q1spin}, that share
the same $p_\varphi^0=3.97$ but have different values of the spins. The horizontal line corresponds to $E_0/M=1.0055$.
The larger centrifugal barrier present when $\chi_1=\chi_2=+0.5$ is responsible of the scattering behavior in
the middle panel of Fig.~\ref{fig:q1spin}. In addition, the case $(\chi_1,\chi_2)=(+0.50,-0.50)$ is extremely close 
to $(\chi_1,\chi_2)=0$ because of spin-orbit coupling cancellation, consistently with the waveform shown in the top panel
of Fig.~\ref{fig:q1spin}.}
\label{fig:spin_potential}
\end{center}
\end{figure}

\begin{figure}[t]
\begin{center}
\includegraphics[width=0.45\textwidth]{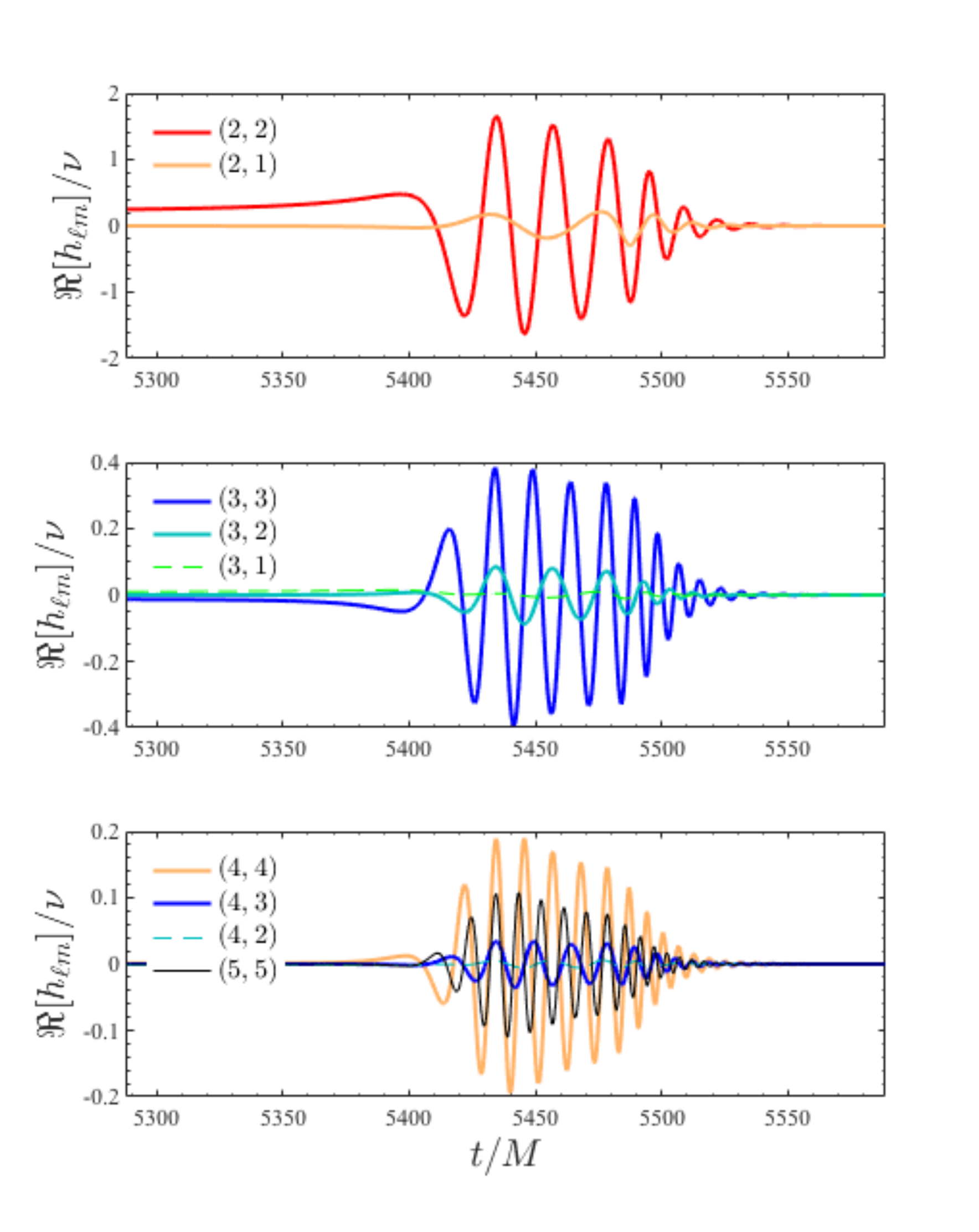}\\
\includegraphics[width=0.4\textwidth]{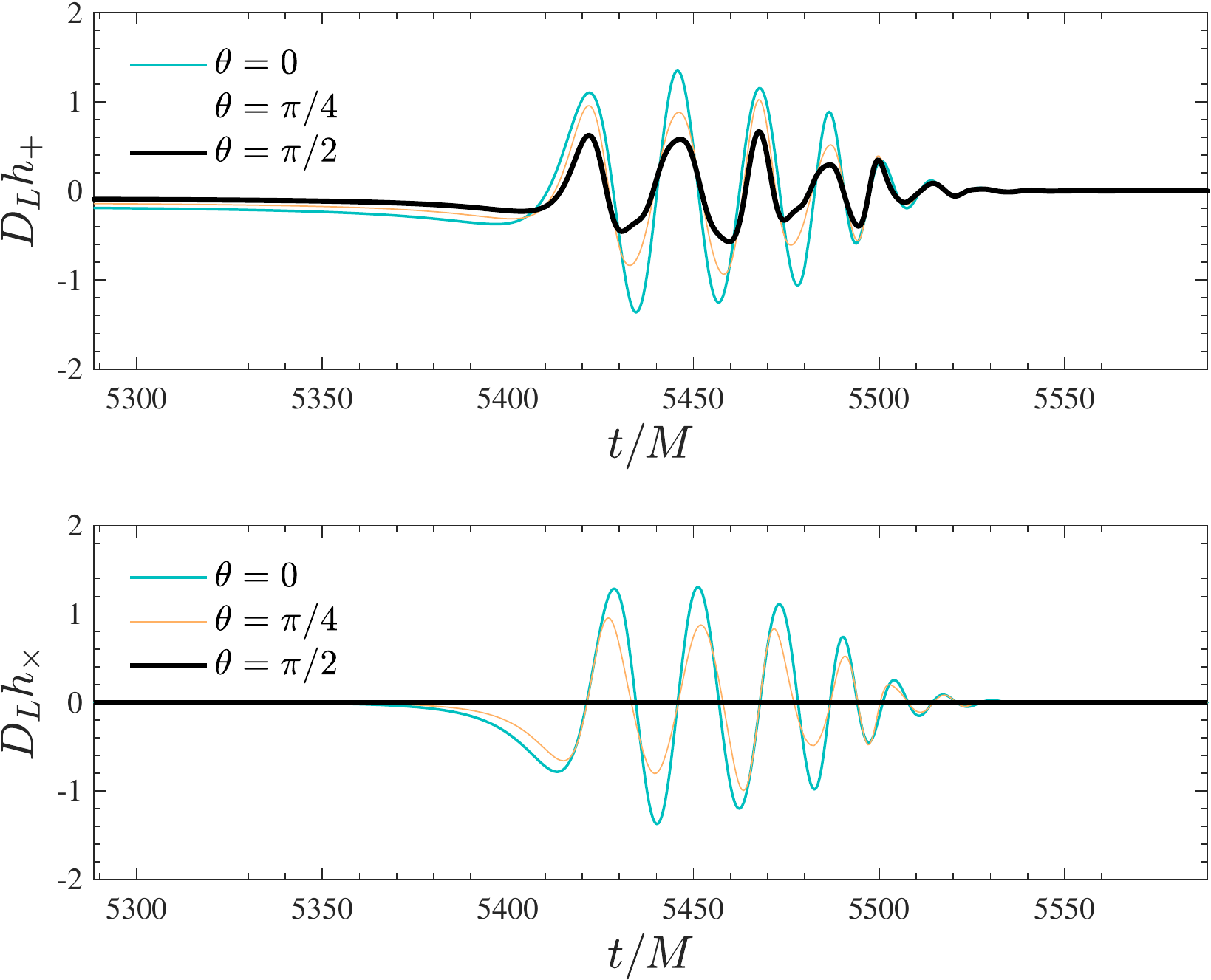}
\caption{Calculation of higher modes for a nonspinning configuration with 
$(q,E_0/M,p_\varphi^0)=(3.5,1.0067,4.1361)$. The phenomenology is qualitatively 
analogous to the $(q,E_0/M,p_\varphi^0)=(1,1.0055,3.97)$ case shown in Fig.~\ref{fig:q1} above.
$D_L h_+$ and $D_L h_\times$ waveform polarizations for $(q,E_0/M,p_\varphi^0)=(3.5,1.0067,4.1361)$ 
at various inclinations: $\theta=0$ (face on), $\theta=\pi/4$ and $\theta=\pi/2$ (edge on). They are obtained 
combining the various multipoles showed using Eq.~\eqref{eq:hpc}.}
\label{fig:q3p5}
\end{center}
\end{figure}
Let us turn now to discussing the effect of the spins (anti)aligned with the angular momentum. At a qualitative
level, the waveform phenomenology is analogous to the nonspinning case considered above, though
with some quantitative differences due to the spin-orbit and spin-spin interactions. 
We shall focus first on a special example to highlight the phenomenology. As mentioned above, 
due to the large initial separation the setup of initial data is insensitive to spin effects, so that the system 
can be consistently started with the same initial data setup for nonspinning binaries discussed above.
In order to single out the effects of spins, we consider the same initial configuration
$(q,E_0/M,p_\varphi^0)=(1,1.0055,3.97)$ as above, with the following three choices for
spins: $\chi_1=\chi_2=+0.50$; $\chi_1=\chi_2=-0.50$; $\chi_1=+0.50$ and $\chi_2=-0.50$. 
The corresponding waveforms are exhibited in Fig.~\ref{fig:q1spin}. One clearly sees the
following facts. When the BHs are spinning in opposite directions, the waveform is essentially 
equivalent to the nonspinning one. This is due to the well known cancellation of the spin-orbit 
interaction in the equal mass case, with the little differences in the waveforms predominantly 
due to spin-spin effects\footnote{Note however that some of the differences in the waveform also 
come from the merger-ringdown modelization, that in one case uses the nonspinning fits, while in the 
other case spin-dependent  fits}. To appreciate this at the level of dynamics, 
Fig.~\ref{fig:spin_potential} shows that the potential energy $H^{\rm EOB}_0/M$ 
(i.e. Eq.~\eqref{eq:H} with $p_{r_*}=0$) for $(\chi_1,\chi_2)=(+0.50,-0.50)$ is 
visually indistinguishable from the nonspinning one. When the spins are both aligned with the orbital 
angular momentum, the centrifugal barrier is higher than in the nonspinning case 
(compare black and red lines in Fig.~\ref{fig:spin_potential}, and thus the system 
undergoes a scattering instead of a capture. The corresponding, burst-like,
waveform is shown in the middle panel of Fig.~\ref{fig:q1spin}.
Finally, when spins are both anti-aligned with the orbital angular momentum, the spin-orbit 
interaction makes the attraction stronger than the nonspinning case (i.e. the potential barrier 
is much lower, see blue curve in Fig.~\ref{fig:spin_potential}) and the system plunges faster,
with a signal whose pre-ringdown phase is much shorter than the nonspinning case.

\subsection{Higher modes}
\label{sec:HM}
Higher modes are incorporated in both the latest quasi-circular 
and eccentric realizations of {\tt TEOBResumS}~\cite{Nagar:2019wds,Nagar:2020pcj,Chiaramello:2020ehz}.
In the nonspinning case, all modes up to $\ell=m=5$ included are robustly completed 
by the NR-informed, quasi-circular, merger and ringdown part~\cite{Nagar:2019wds}. 
By contrast, in the spinning case, due to numerical noise in the NR data, 
it was not possible to model the postmerger-ringdown part in modes modes 
like $(3,1)$, $(4,2)$ and $(4,1)$ (see~\cite{Nagar:2020pcj}).
Figure~\ref{fig:q3p5} shows, in the first three rows, several multipoles 
for $(q,E_0/M,p_\varphi^0)=(3.5,1.0067,4.1361)$. For this choice of initial conditions, 
the system undergoes a quasi-circular orbit before plunge and merger, 
analogously to the corresponding $q=1$ case shown above in the middle panel 
of Fig.~\ref{fig:q1}.  One can appreciate that all modes can be obtained robustly 
with the standard  ringdown matching procedure discussed 
in Refs.~\cite{Nagar:2019wds,Nagar:2020pcj}. 
For visual completeness, the last two rows of the figure also show the corresponding 
two polarizations $(h_+,h_\times)$, obtained using Eq.~\eqref{eq:hpc}, for various 
inclinations.


\section{EOB/NR scattering angle: the equal-mass case}
\label{sec:chi}
\begin{table*}
  \caption{\label{tab:chi_num} EOB scattering angle and comparison with the NR data of Ref.~\cite{Damour:2014afa}.
  These results correspond to the standard configuration of the model, i.e. with 3PN-accurate $D$ and $Q$ functions.
  From left to right, the columns report: the configuration number; the minimum of the EOB radial separation (EOB impact parameter); 
  the initial energy $\hat{E}^{0}$  and angular momentum $p_\varphi^0$; the NR and EOB energy losses; the NR and EOB angular 
  momentum losses; the NR and EOB scattering angles and their fractional difference $\hat{\Delta}\chi^{\rm NREOB}\equiv |\chi^{\rm NR}-\chi^{\rm EOB}|/\chi^{\rm NR}$.
   Angles are measured in degrees. Note that, within the EOB, configuration $\#1$ does not scatter, but plunges instead.}
\begin{ruledtabular}
\begin{tabular}{llllllccllc}
 \# & $r_{\rm min}$ & $\hat{E}^0$ & $p^0_\varphi$ &  $\Delta E^{\rm NR}/M$ & $\Delta E^{\rm EOB}/M$ & $\Delta J^{\rm NR}/M^2$ & $\Delta J^{\rm EOB}/M^2$ &$\chi^{\rm NR}$ & $\chi^{\rm EOB}$  & $\hat{\Delta}\chi^{\rm NREOB}[\%]$\\
 \hline
1& $\dots$ &1.0225555(50)& 4.3986080   & 0.01946(17)       & 0.032553 & 0.17007(89) & 0.363750  & 305.8 (2.6) & $\dots$   &  $\dots$\\
\hline
2&  3.70 &1.0225722(50) & 4.49039348 & 0.01407(10)      & 0.014083&  0.1380(14) & 0.134495 & 253.0(1.4) &   279.35  & 10.4\\
3&  4.03 &1.0225791(50) &4.58209352 &0.010734(75)      & 0.00951037& 0.1164(14) & 0.101919 & 222.9(1.7) &  234.22 & 5.1\\
4&  4.85 &1.0225870(50)& 4.8570920       & 0.005644(38) & 0.0041582& 0.076920(80) & 0.0588254&172.0(1.4) &   174.23 &1.3\\
5&  5.34 & 1.0225870(50)  &  5.0403920  &  0.003995(27)       & 0.00272826  & 0.06163(53)  & 0.045189 &152.0(1.3)  &  153.01 &0.7\\
6&  6.49& 1.0225884(50)   & 5.4986320 &   0.001980(13)  &   0.001172   & 0.04022(53) &  0.027481 & 120.7(1.5) & 120.79 &0.07\\
7&  7.59 & 1.0225924(50) & 5.9568680 & 0.0011337(90) &  0.0005951 & 0.029533(53) & 0.018992 & 101.6(1.7) & 101.51 &0.09\\
8&  8.66& 1.0225931(50) & 6.4150960 & 0.007108(77) &   0.000332568 & 0.02325(47)& 0.0141277    &88.3(1.8) &88.19 &0.12\\
9&  9.72& 1.0225938(50)&  6.8733240  & 0.0004753(75)   & 0.00019778 & 0.01914(76)  & 0.0110359 &78.4(1.8) &  78.28 &0.15\\
10& 10.78 & 1.0225932(50) & 7.33153432 & 0.0003338(77) & 0.0001226  & 0.0162(11) & 0.008928 & 70.7(1.9) &      70.54 & 0.23
\end{tabular}
\end{ruledtabular}
\end{table*}
So far, we have investigated the analytical predictions of our EOB model for dynamical capture 
under the assumption that it provides a reasonably faithful representation of true signals. 
Evidently, seen the many approximations adopted to construct the model, a proof of this statement 
can only come from a systematic analysis of NR simulations of dynamical captures. Unfortunately,
such NR simulations are currently not available. Despite this, we can actually test our model using 
some NR computation of the scattering angle previously published in Ref.~\cite{Damour:2014afa}. 
The scattering angle, $\chi$, is the natural, gauge-invariant, observable that is used to 
characterize hyperbolic encounters. Reference~\cite{Damour:2014afa} provided the first measurement 
of $\chi$ from NR simulations and its comparison with an EOB prediction. The work 
of Ref.~\cite{Damour:2014afa} was a very preliminary investigation of a new territory 
and thus was limited to only $q=1$ binaries. Moreover, the EOB calculation of scattering 
angles of Ref.~\cite{Damour:2014afa} was not EOB-self consistent, since it was relying 
on energy and angular momentum losses computed from NR simulations. 
In this respect, Ref.~\cite{Damour:2014afa} allowed for a detailed analysis
of the properties of the EOB Hamiltonian, but not of the full dynamical model.
Now, thanks to the improved radiation reaction of Ref.~\cite{Chiaramello:2020ehz}, 
reliable in the strong field, we can finally go beyond the approach of~\cite{Damour:2014afa} and 
explore the reliability of the {\it full} model in hyperbolic encounters. This will allow us to
put on a more solid ground the results discussed above.
Reference~\cite{Damour:2014afa} considered 10 configurations, specified by Arnowitt-Deser-Misner (ADM)
energy and angular momentum, of $q=1$ nonspinning black hole binaries. Each configuration
was then evolved numerically. The initial data were chosen so as to always have a scattering 
and not a  capture. Details of the NR simulations are reported in Table~I of~\cite{Damour:2014afa}. 
The values of the dimensionless initial ADM energy $E/M$ and dimensionless initial  angular momentum 
$p_\varphi \equiv J/(M\mu)$ are now listed in the first column of Table~\ref{tab:chi_num}. 
The seventh column of the table collects the values of the NR scattering angle, with their 
uncertainty, as published in Ref.~\cite{Damour:2014afa}. As above, the initial EOB separation 
is chosen to be $r_0=10000$. The EOB values of the scattering angles are listed 
in the eight  column of the table, while the last one lists fractional NR/EOB 
difference, $\hat{\Delta}\chi^{\rm NREOB}\equiv |\chi^{\rm NR}-\chi^{\rm EOB}|/\chi^{\rm NR}$.
A few comments are in order: (i) the EOB/NR agreement between scattering angles is of the 
order of or below $1\%$ fractional difference except for three outliers that correspond, not surprisingly,
to the smallest values of the impact parameter, although such fractional difference is within the NR uncertainty; 
(ii) for the first three configurations, the EOB model systematically overestimates the scattering angle, 
indicating that the system wants to be trapped and eventually plunge, instead of scatter away. 
This is indeed what happens for configuration $\#1$, where the system does a first close encounter, 
followed by a second one and the plunge.

Qualitatively speaking, this behavior is just mirroring the fact that the gravitational attraction 
as modeled within the EOB model is {\it stronger} than the actual NR prediction. At a more 
quantitative level, it is difficult to precisely quantify to which extent this is due to the conservative
or nonconservative part of the dynamics. For what concerns the GW losses, columns 5-8 of 
Table~\ref{tab:chi_num} compare the total fraction of energy and angular momentum 
emitted in the NR simulation with the same quantity computed within the EOB formalism. 
This is what is accounted by the analytical fluxes entering the r.h.s. of Hamilton's equation. 
The analytical fluxes  are seen to always underestimate the numerical ones 
(sometimes also by $\sim 50\%$), except for configuration $\#2$.
Despite this, the estimate of the scattering angle comes out consistent at a few 
percent level up to configuration $\#4$, thus suggesting the crucial importance 
of the conservative  part of the dynamics. We shall come back to this point in
the next section. 

\subsection{Impact of beyond 3PN corrections in the $Q$ and $D$ EOB potentials}
\label{sec:high_PN}

To have a deeper understanding of the results obtained above, let us first remember
that Ref.~\cite{Damour:2014afa} showed that the best EOB/NR agreement was 
obtained by using a $D$ function at (incomplete) 4PN, that was taking into account 
only the linear-in-$\nu$ contributions available at the time~\cite{Bini:2013zaa}.
Now that the 4PN knowledge of the Hamiltonian is complete~\cite{Damour:2015isa,Damour:2016abl},  
the 5PN information is complete except for two undetermined numerical parameters,
$\bar{d}_5^{\nu^2}$ and $a_6^{\nu^2}$ , and similarly the 6PN is known except
for four undetermined numerical 
parameters, $(q_{45}^{\nu^2},\bar{d}_6^{\nu^2},a_7^{\nu^2},a_7^{\nu^3})$~\cite{Bini:2020nsb,Bini:2020hmy},  
it is worth to revive and improve the comparison of Ref.~\cite{Damour:2014afa}. 
Note however that we do this here using the full model with radiation reaction
and taking into account the contributions to either the $D$ and the $Q$ functions
(while Ref.~\cite{Damour:2014afa} was just using the 3PN-accurate $Q$).
In principle, we should also explore, within the present context, the effect of higher 
PN corrections to the $A$ function. However, we decided not to do so now for the
following two reasons. On the one hand, the analytically known numerical value of 
the 5PN correction to the potential $a_6^c$  is such that the usual $(1,5)$ 
Pad\'e approximant has a spurious pole, making thus this additional analytical 
knowledge practically useless within the current EOB context. Exploring different 
resummation strategies (e.g. changing Pad\'e approximant) would be necessary 
in order to fruitfully use the analytically known 5PN result. On the other hand,
we have verified that even large changes ($\sim 100\%$) of the NR-informed 
effective 5PN parameter $a_6^c(\nu)$ obtained in Refs.~\cite{Nagar:2019wds,Nagar:2020pcj}
and used here have little to negligible impact on the calculation of the scattering
angle within the EOB model. This is consistent with the fact that the $A$ function
rules the azimuthal part of the energy and it is less important in a hyperbolic-like
context when the radial part of the Hamiltonian, i.e. $\propto p_{r_*}^2$ becomes
predominant. To avoid additional complications we thus prefer to keep working
with the NR-informed expression of the $A$ function used in previous work, 
focusing instead only on the high-PN corrections to the $Q$ and $D$ functions.

\subsubsection{Q function}
Let us start discussing the $Q$ function. To simplify the logic, we keep $D$ fixed at
3PN order and consider only $Q$ at 4PN and at 5PN, though incorporating only local
terms. The result of the $\chi$ computation is displayed in Table~\ref{tab:Q}. One sees that the 4PN
terms bring a small, though significative, contribution to the scattering angle that goes
in the direction of reducing the EOB/NR difference. Despite this, the magnitude of the correction
is too small to avoid configuration $\#1$ to plunge. By contrast, the effect of the 5PN 
local-in-time terms goes in the wrong direction and, moreover, is significantly smaller than
the numerical uncertainty. At a practical level, and especially in view of the analytic complexity
of the $Q$ function at 6PN, we don't think it is worth, for the current study, to push $Q$ at 6PN accuracy
and we shall just work, from now on, at 4PN accuracy in $Q$.
\begin{table}
		\caption{\label{tab:Q}Impact of 4PN and 5PN terms in the $Q$ function on
		the calculation of the scattering angle $\chi$. Angles are measured in degrees}
	\begin{ruledtabular}
		\begin{tabular}{ccccc}
			\#  &$\chi^{\rm NR}$ & $\chi^{\rm EOB}_{Q_{\rm 3PN}}$ & $\chi^{\rm EOB}_{Q_{\rm 4PN}}$ & $\chi^{\rm EOB}_{Q_{\rm 5PN}}$ \\
			\hline 
			1 & 305.8(2.6)   &  $\dots$    & $\dots$  \\
    			2 & 253.0(1.4)   &  279.35     & 278.21 &  278.75 \\
   			3 & 222.9(1.7)   &   234.22    & 233.27 &   233.62 \\
                         4 & 172.0(1.4)    &  174.23    & 173.57 &   173.72 \\
    			5 &  152.0(1.3)   &  153.01    & 152.47 &   152.57 \\
    			6 &  120.7(1.5)   &  120.79    & 120.44 &     120.49 \\
    			7 &  101.6(1.7)   &  101.51    & 101.28 &    101.29\\
   			8 &   88.3(1.8)    &  88.19      & 88.03 &      88.04 \\
    			9 &   78.4(1.8)    &  78.28      & 78.16 &     78.17 \\
  			10 &  70.7(1.9)   &  70.54      & 70.44 &   70.45 
		\end{tabular}
	\end{ruledtabular}
\end{table}

\begin{figure}[t]
\begin{center}
\includegraphics[width=0.45\textwidth]{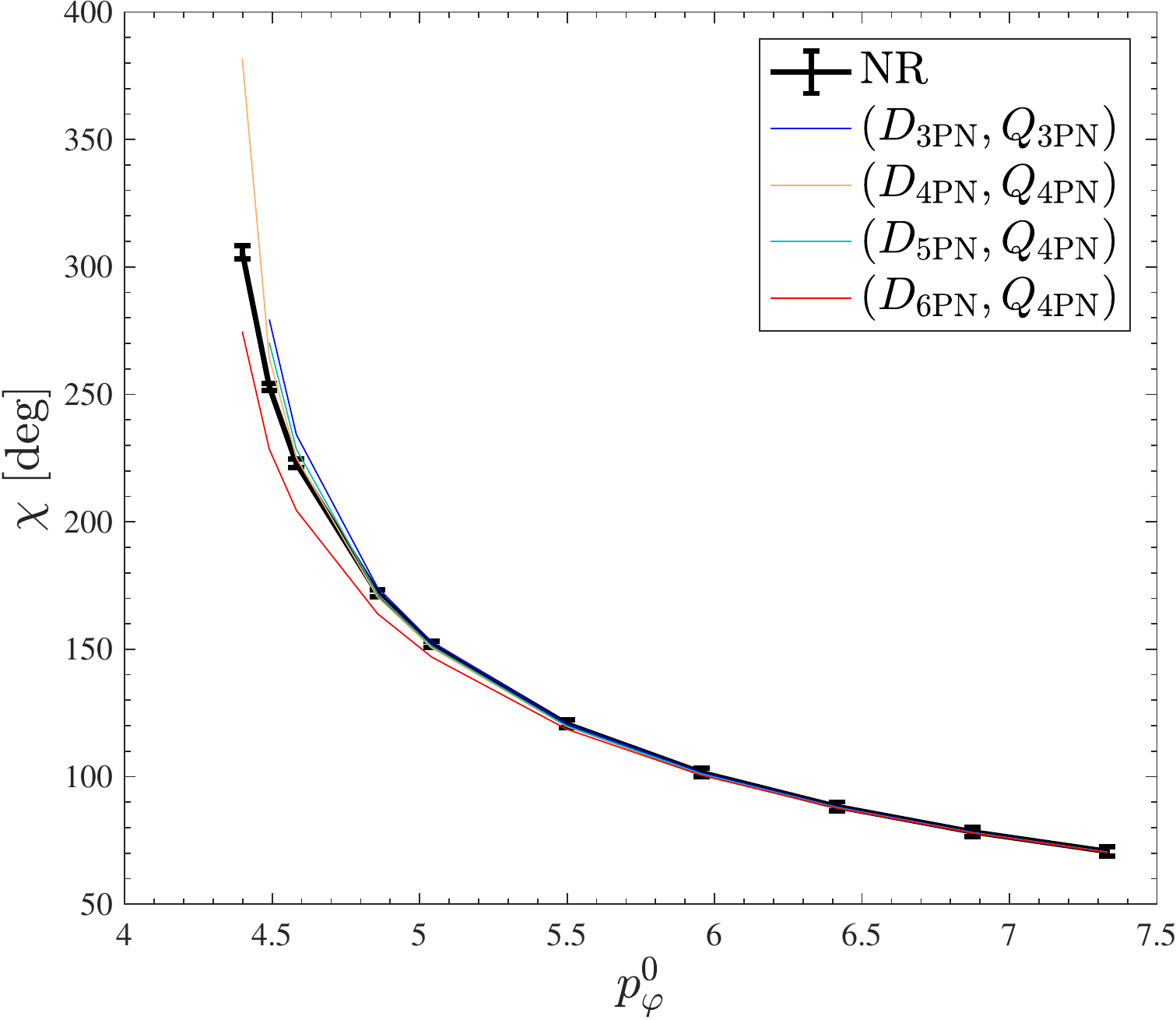}
\caption{Comparing the NR scattering angles with various EOB predictions using different 
PN accuracies of the $(Q,D)$ potential. The 6PN-accurate $D$ function allows for the closest 
EOB/NR agreement for the smallest values of the EOB impact parameter.}
\label{fig:chi}
\end{center}
\end{figure}
\begin{table}
\caption{\label{tab:D4PN_Q4PN}EOB scattering angle obtained using both $D$ and $Q$ functions and 4PN. 
The EOB/NR agreement is improved with respect to the standard case of Table~\ref{tab:chi_num} that adopt
3PN accuracy for these functions. Note that configuration $\#1$ does not plunge anymore.}
\begin{ruledtabular}
\begin{tabular}{ccccccc}
 \# & $r_{\rm min}$ & $\Delta E^{\rm EOB}/M$ & $\Delta J^{\rm EOB}/M^2$ &$\chi^{\rm NR}$ & $\chi^{\rm EOB}$  & $\hat{\Delta}\chi~[\%]$\\
\hline
1   &3.31     &0.022693    & 0.190585   &305.8   &381.93  &   24.89 \\
 2  & 3.71    &0.012995     &0.126256   &253.0   &264.21  &   4.43  \\
3   & 4.03     &0.008920     &0.097128   &222.9   &225.12  &   0.99  \\
4   & 4.85     &0.003997     &0.057269   &172.0   &170.53  &   0.85  \\
5   & 5.34     &0.002646    & 0.044311   &152.0   &150.60  &   0.92  \\
6   & 6.49     &0.001151     &0.027202   &120.7   &119.72  &   0.81  \\
7    &7.59    & 0.000588     &0.018878   &101.6  & 100.93  &   0.66  \\
8   & 8.66    & 0.000330     &0.014074    &88.3  &  87.85   &  0.51  \\
9    &9.72    & 0.000196     &0.011008    &78.4 &    78.05   &  0.44 \\
10   &10.78 &    0.000122    & 0.008912   & 70.7&    70.38  &   0.45 
\end{tabular}
\end{ruledtabular}
\end{table}
\begin{table}
\caption{\label{tab:D5PN_Q4PN}EOB scattering angle computed using $Q$ at 4PN and 
$D$ at 5PN, though resummed with a $(1,4)$ Pad\'e approximant since $P^0_5$ develops
a spurious pole.The EOB/NR agreement is worsened with respect to Table~\ref{tab:D4PN_Q4PN} 
and configuration $\#1$ plunges again. See text for discussion.}
\begin{ruledtabular}
\begin{tabular}{ccccccc}
 \# & $r_{\rm min}$ & $\Delta E^{\rm EOB}/M$ & $\Delta J^{\rm EOB}/M^2$ &$\chi^{\rm NR}$ & $\chi^{\rm EOB}$  & $\hat{\Delta}\chi~[\%]$\\
\hline
1   &  $\dots$   &   0.031705   &  0.347309  & 305.8  & plunge &    $\dots$ \\
2   & 3.71  &   0.013463   &  0.129751  & 253.0  & 270.26   &  6.82 \\
3   & 4.03   &  0.009161   &  0.099057  & 222.9  & 228.48   &  2.50 \\
4   & 4.85   &  0.004054   &  0.057809  & 172.0  & 171.63   &  0.22 \\
5   & 5.34    & 0.002673   &  0.044589  & 152.0  & 151.22   &  0.51 \\
6   & 6.49    & 0.001157   &  0.027273  & 120.7  & 119.92   &  0.65 \\
7   & 7.59    & 0.000590  &   0.018901  & 101.6  & 101.01   &  0.58 \\
8   & 8.66    & 0.000330   &  0.014083  &  88.3   & 87.88   &   0.47 \\
9   & 9.72    & 0.000197   &  0.011012  &  78.4   & 78.07   &   0.42 \\
10 &  10.78  &   0.000122  &   0.008914 &   70.7  &  70.39 &    0.44
\end{tabular}
\end{ruledtabular}
\end{table}
\begin{table}
\caption{\label{tab:D6PN_Q4PN}EOB scattering angle computed using $D$ function at 6PN
		and the $Q$ function at 4PN~\cite{Bini:2020wpo,Bini:2020nsb}. The 6PN-accurate $D$ function is
		essential to get an improved EOB/NR agreement for small values of the EOB impact
		parameter. Note, however, that this also brings slightly larger deviations with respect to the 
		previous cases for intermediate values of $r_{\rm min}$.}
\begin{ruledtabular}
	\begin{tabular}{ccccccl}
		\# & $r_{\rm min}$ & $\Delta E^{\rm EOB}/M$ & $\Delta J^{\rm EOB}/M^2$ &$\chi^{\rm NR}$ & $\chi^{\rm EOB}$  & $\hat{\Delta}\chi~[\%]$\\
			\hline

1  &  3.33   &  0.015559   &  0.141465  & 305.8  &  274.68  &  10.18 \\
2  &  3.71   &  0.010137   &  0.105088  & 253.0  &  228.49  &   9.69 \\
3  &  4.03   &  0.007422   &  0.085263  & 222.9  &  204.52  &   8.24 \\
4  &  4.85   &  0.003654   &  0.054047  & 172.0  &  163.99  &   4.66 \\
5  &  5.34   &  0.002490   &  0.042707  & 152.0  &  146.99  &   3.30 \\
6  &  6.49   &  0.001121   &  0.026816  & 120.7  &  118.63  &   1.71 \\
7  &  7.59   &  0.000580   &  0.018755  & 101.6  &  100.51  &   1.07 \\
8  &  8.66   &  0.000327   &  0.014027  &  88.3  &   87.66  &   0.72 \\
9  &  9.72   &  0.000195   &  0.010987  &  78.4  &   77.96  &   0.56 \\
10 & 10.78   &  0.000122   &  0.008902  &  70.7  &   70.33  &   0.52

		\end{tabular}
	\end{ruledtabular}
\end{table}

\subsubsection{D function}
Now that we have explored the (ir)relevance of the various PN truncations of the $Q$
function, let us move to exploring the $D$ function. We consider all terms up to 6PN,
i.e. separately work with 4PN, 5PN and 6PN truncations, keeping the accuracy
of $Q$ fixed at 4PN. Each $D$ function, that comes as a PN-truncated series, is resummed.
Let us write here, for completeness, the Taylor expansion of $D$ up to 6PN
\begin{align}
\label{eq:D}
D^{\rm Taylor}_{\rm 6PN}&= 1 - 6\nu u^2 - (52\nu - 6\nu^2)u^3 + \nu d_4 u^4 \nonumber\\
                      &+ \nu d_5 u^5 + \nu d_{5.5} u^{11/2} + \nu d_6 u^6,
\end{align}
The standard resummation procedure for this function is to take a $P^0_n$ approximant
of this equation. This approach is now so standard that any analytical result is usually 
given in terms of the denominator, $\bar{D}$ function, where $\bar{D}\equiv 1/D$
(see e.g. Refs.~\cite{Bini:2019nra,Bini:2020wpo,Bini:2020nsb,Bini:2020hmy}). 
The coefficients $(d_4,d_5,d_{5.5},d_6)$ in Eq.~\ref{eq:D} are obtained by just expanding 
$1/\bar{D}$ as given in the literature. One finds that the 5PN $D$ function resummed taking the $(0,5)$ 
Pad\'e approximant has a spurious pole around $u\approx 0.5$ and thus it cannot be used 
robustly to deliver analytical prediction. That is the reason why we prefer
to give the results of Refs.~\cite{Bini:2019nra,Bini:2020wpo,Bini:2020nsb,Bini:2020hmy}
in terms of the $D$ function in Eq.~\eqref{eq:D} and then, at 5PN accuracy, proceed by
resumming it with a $(1,4)$ Pad\'e approximant, that is $D_{\rm 5PN}=P^1_4\left[D^{\rm Taylor}_{\rm 5PN}\right]$, 
that is found to have a pole-free behavior.
The results of the calculations of the scattering angle with higher PN knowledge in $D$ 
are listed in Tables~\ref{tab:D4PN_Q4PN}-\ref{tab:D6PN_Q4PN}.
The following conclusions are in order: (i) increasing the analytic information of $D$ to 4PN
brings the first, important, qualitative and quantitative improvement, since configuration $\#1$
is found to correctly scatter (though the scattering angle is still significantly smaller than the NR one) 
instead of plunging; (ii) moving to 5PN is a step back, since configuration $\#1$ plunges again.
By contrast, (iii), a remarkable improvement is obtained working at 6PN, retaining
all the currently unknown numerical parameters fixed to zero. For the smallest impact parameters, 
the EOB/NR difference is at most of the order of $10\%$, an improvement of more than a factor 
two with respect to the cases discussed above.
One also notes, however, that for intermediate values of the EOB impact parameter the EOB/NR 
agreement is slightly less good than, for instance, the $(D_{\rm 4PN},Q_{\rm 4PN})$ case.
\begin{figure}[t]
\begin{center}
\includegraphics[width=0.5\textwidth]{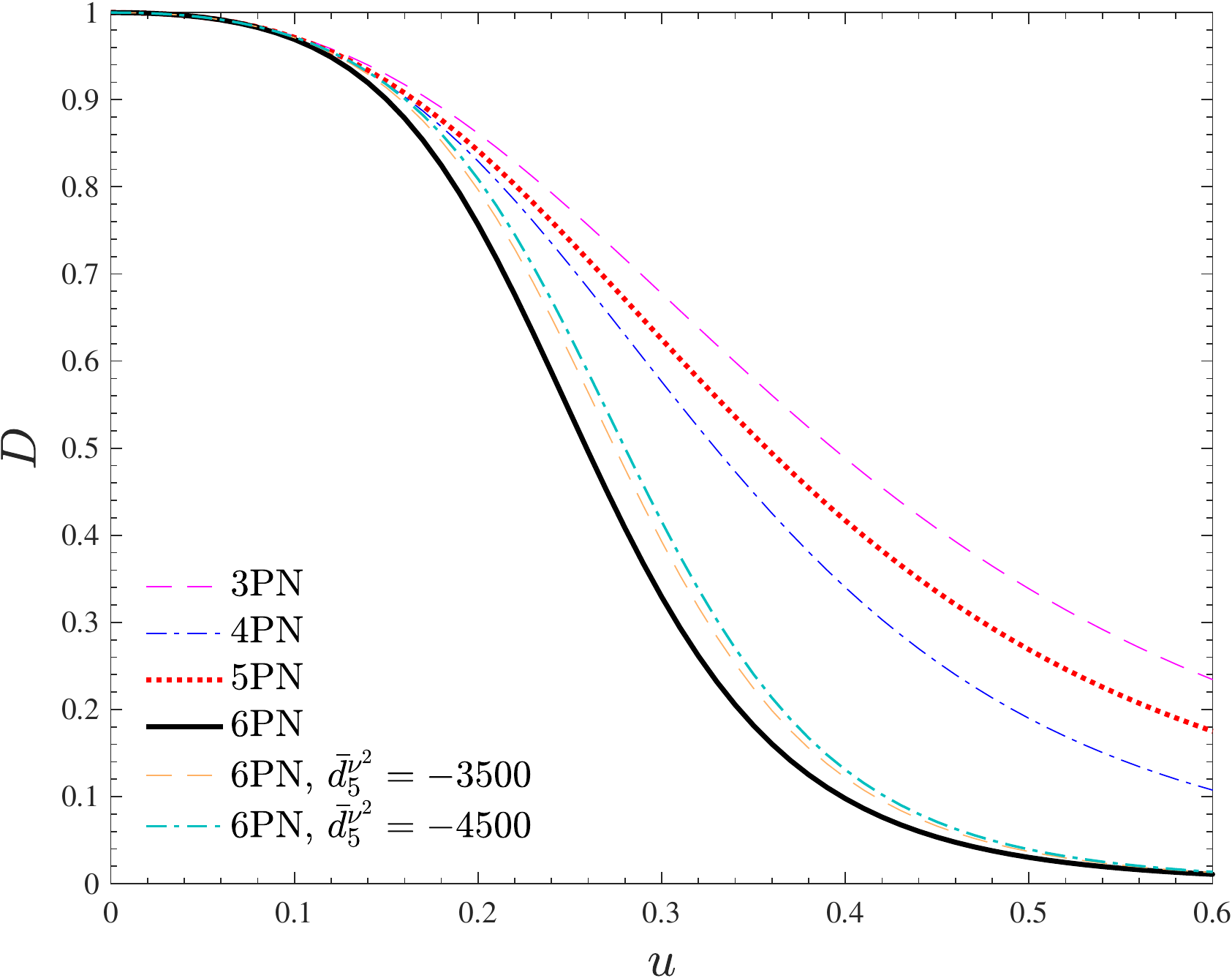}
\caption{The resummed $D$ function for $q=1$ at various PN orders. The function gets progressively
steeper as the PN order is increased. The orange lines corresponds to the {\it flexed} 6PN function where
we put $d_5^{\nu^2}=-3500$ (dashed) and $d_5^{\nu^2}=-4500$ (dash-dotted). This latter brings the 
EOB/NR agreement of the order of percent for any configuration considered. See Table~\ref{tab:6PN_tuned}
and additional discussion in text.}
\label{fig:Dfun}
\end{center}
\end{figure}
\begin{table}
		\caption{\label{tab:6PN_tuned}Tuned $D$ potential with $\bar{d}_5^{\nu^2}=-3500$.
		This number is chosen so to have an excellent EOB/NR agreement for configuration $\#1$.
		Still, for half of the configurations the EOB/NR difference is slightly larger than the NR error bar.
		See text for additional discussion.
		}
	\begin{ruledtabular}
		\begin{tabular}{ccccccl}
			\# & $r_{\rm min}$ & $\Delta E^{\rm EOB}/M$ & $\Delta J^{\rm EOB}/M^2$ &$\chi^{\rm NR}$ & $\chi^{\rm EOB}$  & $\hat{\Delta}\chi[\%]$\\
			\hline
1 &3.32  &   0.017860  &   0.157308  & 305.8  & 303.17   &  0.86 \\
2 &3.71   &  0.011289   &  0.113663  & 253.0  & 243.02   &  3.9 \\
3 &4.03  &   0.008109  &   0.090735 &  222.9  & 214.15  &   3.9 \\
4 &4.85   &  0.003853  &  0.055929  & 172.0  & 167.87   &  2.4 \\
5 &5.34   &  0.002591   &  0.043749  & 152.0  & 149.36  &   1.7 \\
6 & 6.49  &   0.001145   &  0.027120 & 120.7  & 119.50  &   0.10 \\
7 & 7.59  &   0.000587   &  0.018867  & 101.6  & 100.89  &   0.69 \\
8 & 8.66  &   0.000330   &  0.014075  &  88.3  &  87.85    &  0.51 \\
9 & 9.72   &  0.000197   &  0.011010  &  78.4   & 78.06    &  0.43 \\
10 &10.78  &   0.000122  &   0.008914  &  70.7  &  70.39  &   0.44
		\end{tabular}
	\end{ruledtabular}
\end{table}
It is useful to visualize the functional behavior of the resummed $D$ function for the various PN
orders considered above, see Fig.~\ref{fig:Dfun}. The effect of higher PN order is to reduce the
magnitude of the $D$ potential for $u\gtrsim 0.15$, i.e. $r\lesssim 7$. This is indeed the regime
of radii explored  by configurations $\#1-\#7$. The figure then indicates that the  $D$ function that
best approximates the NR values of the scattering angle (within the current analytical framework)
should be slightly larger than the analytically known 6PN one, starting from $u\approx 0.18$.
Still, it has to stay well below the 4PN curve.
Although our finding is rather interesting because it demonstrates that, by varying a single 
analytical element, one can progressively improve the EOB/NR agreement of the scattering 
angle, it is not yet satisfactory because the difference is still larger than the NR error bar.
It is then reasonable to ask whether it is possible to effectively {\it flex}  the current
$D_{\rm 6PN}$ so as to further improve the EOB/NR agreement for the smallest 
values of $r_{\rm min}$. As noted above $D_{\rm 6PN}$ is analytically known modulo 
three parameters, $(a_6^{\nu^2},\bar{d}_5^{\nu^2},\bar{d}_6^{\nu^2})$. 
We found that changing only $\bar{d}_5^{\nu^2}$ gives us enough flexibility for 
our aim. Figure~\ref{fig:Dfun} exhibits, with a orange line, the curve  corresponding 
to $\bar{d}_5^{\nu^2}=-3500$. This value of the parameter was determined so to
provide an EOB/NR agreement below the percent level for the smallest values 
of the impact parameter, as shown in Table~\ref{tab:6PN_tuned}. One should 
however note that for half of the configurations, the EOB/NR difference is still 
{\it larger}  than the NR error bar, that is always of the order  of percent or smaller. 
One then verifies that $\bar{d}_5^{\nu^2}=-4500$ allows one to obtain values of
$\hat{\Delta}\chi=(2.52,2.07,2.57,1.73,1.13)\%$ for the first five configurations, 
and below $1\%$ for the following ones, i.e. $\hat{\Delta}\chi=(0.79,0.59,0.45,0.39,0.41)\%$.
One should note, however, that $\chi_{\#1}^{\rm EOB}=313.51$, i.e. it is 
now {\it larger} than the NR value. The curve with $\bar{d}_5^{\nu^2}=-4500$ 
is also shown on Fig.~\ref{fig:Dfun} for completeness.
Evidently, seen the still large errors in the NR computations, that date back 
to a few years ago, the various approximations involved in our analytical model
(notably, the radiation reaction) and various possibilities of tuning free parameters,
we do not want to make any strong claim about the physical meaning to the
NR-tuning of $\bar{d}_5^{\nu^2}$. Still, our exercise shows that there is a large
amount of yet unexplored analytical flexibility within the EOB model that can be 
constrained using the NR knowledge of the scattering angle, as originally 
advocated in Ref.~\cite{Damour:2014afa}. 

\begin{figure}[t]
\begin{center}
\includegraphics[width=0.45\textwidth]{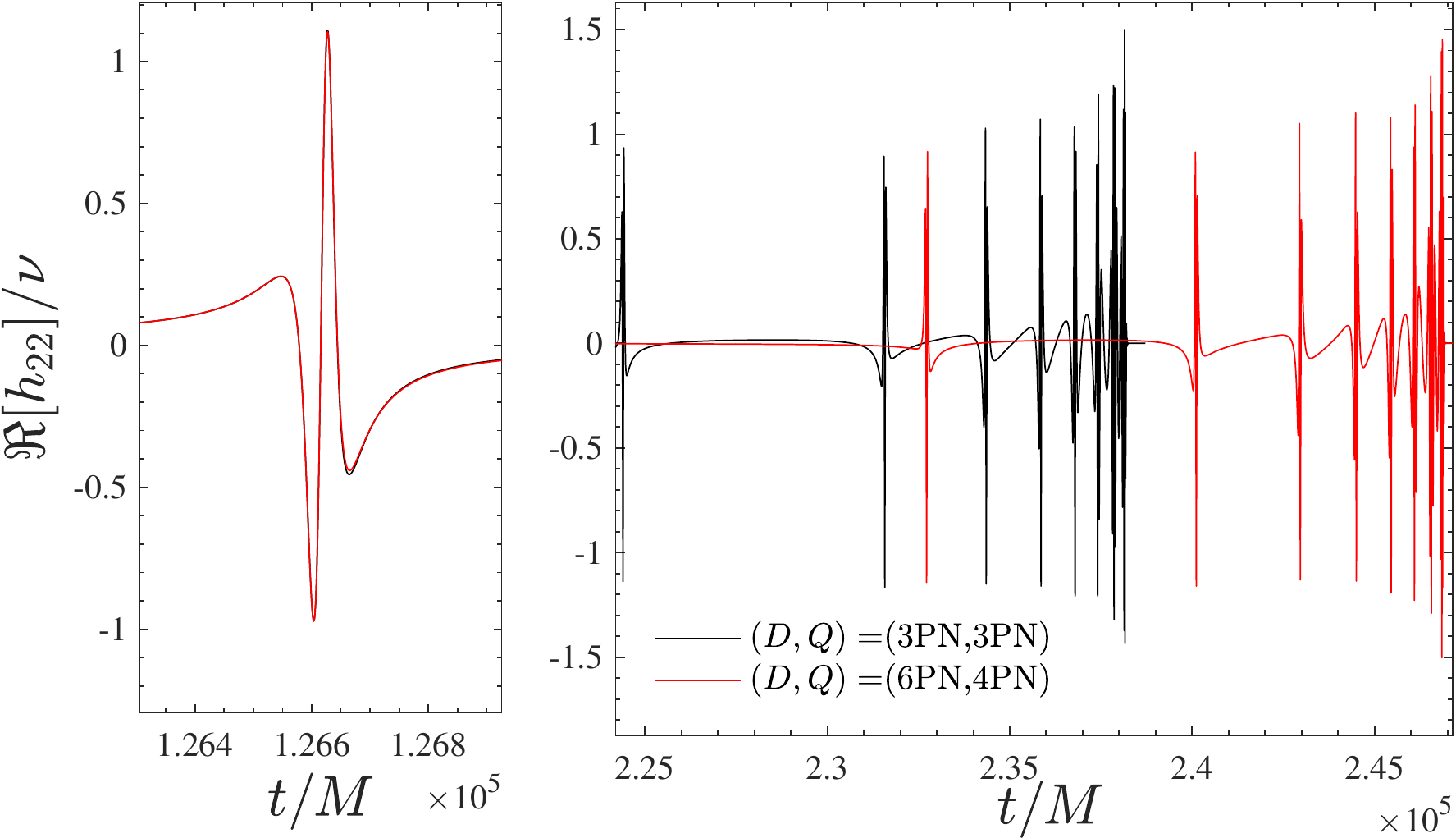}
\caption{Waveforms for configuration with $(q,E_0/M,p_\varphi^0)=(8,1.0003,4.3142)$ considered 
in Figs.~\ref{fig:q8config} and~\ref{fig:q8_geo} with $(D,Q)$ at 3PN (resummed) accuracy contrasted 
with the analytical prediction obtained with $D$ at 6PN and $Q$ at 4PN. The left panel compares the
waveforms during the first encounter; the right panel the subsequent periastron passages, up to the final
merger. The phenomenology is qualitatively the same, but the less attractive character of $D_{\rm 6PN}$ 
with respect to $D_{\rm 3PN}$  results in a larger time-lag between one encounter and the other and in 
a waveform that is globally almost twice longer.}
\label{fig:q8_3pn6pn}
\end{center}
\end{figure}
As an additional exploratory study, we show in Fig.~\ref{fig:q8_3pn6pn} how would change the waveform for
$(q,E_0/M,p_\varphi^0)=(8,1.0003,4.3142)$ of Fig.~\ref{fig:q8config} when we replace
$(D_{\rm 3PN},Q_{\rm 3PN})$ with $(D_{\rm 6PN},Q_{\rm 4PN})$. The less attractive character
of $D_{\rm 6PN}$, as discussed above, results in a larger time-lag between the various bursts
after the first encounter and the waveform is almost twice longer than with $D_{\rm 3PN}$ case.
We conclude that there is a urgent need of specifically tuned NR simulations of dynamical capture
black hole binaries aiming at understanding to which extent the analytical elements entering our
model are trustable and what needs to be changed in order to achieve a level of NR-faithfulness
sufficient for parameter-estimation purposes.

\section{Conclusions}
\label{sec:conclusions}

We presented an EOB model to describe the dynamics of 
spin-aligned BBH hyperbolic encounters and the emitted gravitational waveform.
The model generalizes the most developed version of {\tt TEOBResumS} (named {\tt TEOBiResumS\_SM}) 
to dynamical captures, in particular building on the approach introduced in Ref.~\cite{Chiaramello:2020ehz}. 
The dynamics includes radiation reaction and multipolar waveforms for BBH with arbitrary mass ratio and 
aligned-spin interactions. Our main findings can be summarized as follows:
\begin{itemize}
\item[(i)] We have extensively explored the parameter space of nonspinning dynamical captures,
parameterizing it in terms of initial energy and angular momentum. In particular, we have 
characterized various regions on the basis of the number of close encounters that happen 
before merger, that are measured looking at the number of peak of the orbital frequency.
We have found that the region of parameter space with two peaks, i.e. an encounter followed
by the merger, gets smaller and smaller as the mass ratio increases. By contrast, the number
of encounters before merger in the special region close to the stability regime increases
with the mass ratio. The dynamical behavior mirrors in the waveform, that is completed with 
a merger and ringdown part informed by quasi-circular NR-simulation. The waveform incorporates 
multipoles up to $\ell=m=5$. Modes with $m=0$ are currently missing in the model.
\item[(ii)] We have briefly explored the effect of spin, in order to get a qualitative idea of the general
behavior. The effect of spins is essentially two fold. When spin are aligned with respect to
the orbital angular momentum, a capture that is present in the nonspinning case, may transform
in scattering if the spin-orbit interaction is sufficiently strong. By contrast, spins anti-aligned
with the orbital angular momentum accelerate the capture process, so that the corresponding
waveform eventually ends up with less gravitational wave cycles, and is dominated by the final 
ringdown part. 
\item[(iii)] Beyond the need, for GW-data purposes, of providing an analytical description of
the dynamics and radiation of relativistic hyperbolic encounters, we have refreshed the EOB/NR
comparison between scattering angles $\chi$ that was pioneered in Ref.~\cite{Damour:2014afa}.
Two are our most relevant findings. On the one hand, working with the $D$ function at 3PN, 
i.e. exactly with the eccentric EOB model of Ref.~\cite{Chiaramello:2020ehz}, we showed that
an EOB-self-consistent calculation of the scattering angle (that is, including radiation reaction)
is well compatible with the NR results of Ref.~\cite{Damour:2014afa}, although things become
quantitatively and qualitatively different (i.e. plunge instead of scattering) for the smallest value
of the EOB  impact parameter.

On the other hand, we have systematically explored the impact of 4PN, 5PN and 6PN corrections
to the $D$ function. Our most important finding is that the recently computed, 6PN-accurate, 
$D$ function allows one to obtain an EOB/NR agreement for the scattering angle of a few 
percents {\it also} for the configurations with the smallest impact parameter. 
We thus argue that our model for hyperbolic scattering/dynamical
capture is, probably, more accurate using $D_{\rm 6PN}$ instead of $D_{\rm 3PN}$. Still, one has to mention
that the improvement for small values of $r_{\rm min}$ is balanced by a slight worsening of the $\chi$
computation for intermediate values of the impact parameter. Due to the absence of additional 
NR simulations all over the parameter space, we take the difference between $D_{\rm 3PN}$ 
and $D_{\rm 6PN}$ results as a (rather conservative) error bar that might be taken into account, 
using the current model, in a possible parameter estimation on a GW detection 
qualitatively and morphologically compatible with a dynamical capture scenario.

We have also shown that it is rather easy to additionally tune the $D_{\rm 6PN}$ 
function so as to further improve the EOB/NR agreement of the scattering angle, at the level of
the actual estimate of the uncertainty on the NR scattering angle. This is done by tuning only
the uncalculated 5PN numerical parameter $\bar{d}_5^{\nu^2}$. Concretely, our results 
indicate that, one the one hand, it would be a good idea to  incorporate the 6PN-accurate $D$ function 
in waveform model for quasi-circular coalescing BBHs; on the other hand, it proves that NR 
simulations of the scattering angle can be used to {\it inform} the EOB model and thus 
dedicated NR simulations to systematically and usefully explore the parameter space should 
be performed at some stage. Similarly, systematic NR surveys of dynamical capture are needed
to test our model and to possibly improve the merger-ringdown part, that at the moment is
informed by quasi-circular simulations. Also, dedicated simulations in the large-mass-ratio limit,
e.g. using {\tt Teukode}~\cite{Harms:2014dqa}, would be needed to provide strong tests of the
model at a moderate computational cost. This investigation will be pursued in future work.
This strategy seems the only viable at the moment in order to complete a multi-purpose
EOB-based waveform model able to take into account any kind of coalescence configuration,
from quasi-circular ones, to eccentric up to dynamical capture.
\end{itemize}
Our waveform model is implemented as a standalone $C$ code that is publicly 
available via a {\tt bitbucket} repository~\cite{teobresums}. Details of the implementation 
are also discussed in Ref.~\cite{Chiaramello:2020ehz}. Its performance for parameter 
estimation and optimization is under way and will be reported in a separate publication.
We finally note that the recently published GW signal GW190521~\cite{Abbott:2020tfl,Abbott:2020mjq},
though interpreted as a BBH coalescence with total mass $M\simeq 142M_\odot$ using
precessing, quasi-circular, templates, has a morphology compatible with  that of 
a dynamical  capture. As advocated in Ref.~\cite{CalderonBustillo:2020odh}
(see also~\cite{Romero-Shaw:2020thy,Gayathri:2020coq}), our model could be used in 
the future to attempt to rule out that GW190521 is actually the result of a dynamical 
capture BBH merger.

\begin{acknowledgments}
 We thank T.~Damour for comments on the manuscript and T.~Font for discussions.
 We also warmly thank D.~Chiaramello for collaboration in the very early stages of
 this work. We are also grateful to M.~Breschi, for valuable help in the implementation,
 and to I.~Romero-Shaw for constructive comments and criticisms on an earlier version
 of the manuscript.
  R.~G. acknowledges support from the Deutsche Forschungsgemeinschaft
  (DFG) under Grant No. 406116891 within the Research Training Group
  RTG 2522/1. 
  M.~B., S.~B. acknowledge support by the EU H2020 under ERC Starting
  Grant, no.~BinGraSp-714626.
  M.~B. acknowledges partial support from the Deutsche Forschungsgemeinschaft
  (DFG) under Grant No. 406116891 within the Research Training Group
  RTG 2522/1. 
  The computational experiments were performed on resources of Friedrich
  Schiller University Jena supported in part by DFG grants INST 275/334-1 FUGG and INST 275/363-1 FUGG
  and on the {\tt Tullio} INFN cluster at INFN Turin.
\end{acknowledgments}

\appendix

\bibliography{refs20200927.bib,local.bib}

\begin{thebibliography}{44}%
\makeatletter
\providecommand \@ifxundefined [1]{%
 \@ifx{#1\undefined}
}%
\providecommand \@ifnum [1]{%
 \ifnum #1\expandafter \@firstoftwo
 \else \expandafter \@secondoftwo
 \fi
}%
\providecommand \@ifx [1]{%
 \ifx #1\expandafter \@firstoftwo
 \else \expandafter \@secondoftwo
 \fi
}%
\providecommand \natexlab [1]{#1}%
\providecommand \enquote  [1]{``#1''}%
\providecommand \bibnamefont  [1]{#1}%
\providecommand \bibfnamefont [1]{#1}%
\providecommand \citenamefont [1]{#1}%
\providecommand \href@noop [0]{\@secondoftwo}%
\providecommand \href [0]{\begingroup \@sanitize@url \@href}%
\providecommand \@href[1]{\@@startlink{#1}\@@href}%
\providecommand \@@href[1]{\endgroup#1\@@endlink}%
\providecommand \@sanitize@url [0]{\catcode `\\12\catcode `\$12\catcode
  `\&12\catcode `\#12\catcode `\^12\catcode `\_12\catcode `\%12\relax}%
\providecommand \@@startlink[1]{}%
\providecommand \@@endlink[0]{}%
\providecommand \url  [0]{\begingroup\@sanitize@url \@url }%
\providecommand \@url [1]{\endgroup\@href {#1}{\urlprefix }}%
\providecommand \urlprefix  [0]{URL }%
\providecommand \Eprint [0]{\href }%
\providecommand \doibase [0]{http://dx.doi.org/}%
\providecommand \selectlanguage [0]{\@gobble}%
\providecommand \bibinfo  [0]{\@secondoftwo}%
\providecommand \bibfield  [0]{\@secondoftwo}%
\providecommand \translation [1]{[#1]}%
\providecommand \BibitemOpen [0]{}%
\providecommand \bibitemStop [0]{}%
\providecommand \bibitemNoStop [0]{.\EOS\space}%
\providecommand \EOS [0]{\spacefactor3000\relax}%
\providecommand \BibitemShut  [1]{\csname bibitem#1\endcsname}%
\let\auto@bib@innerbib\@empty
\bibitem [{\citenamefont {Rasskazov}\ and\ \citenamefont
  {Kocsis}(2019)}]{Rasskazov:2019gjw}%
  \BibitemOpen
  \bibfield  {author} {\bibinfo {author} {\bibfnamefont {A.}~\bibnamefont
  {Rasskazov}}\ and\ \bibinfo {author} {\bibfnamefont {B.}~\bibnamefont
  {Kocsis}},\ }\href {\doibase 10.3847/1538-4357/ab2c74} {\bibfield  {journal}
  {\bibinfo  {journal} {Astrophys. J.}\ }\textbf {\bibinfo {volume} {881}},\
  \bibinfo {pages} {20} (\bibinfo {year} {2019})},\ \Eprint
  {http://arxiv.org/abs/1902.03242} {arXiv:1902.03242 [astro-ph.HE]}
  \BibitemShut {NoStop}%
\bibitem [{\citenamefont {Tagawa}\ \emph {et~al.}(2020)\citenamefont {Tagawa},
  \citenamefont {Haiman},\ and\ \citenamefont {Kocsis}}]{Tagawa:2019osr}%
  \BibitemOpen
  \bibfield  {author} {\bibinfo {author} {\bibfnamefont {H.}~\bibnamefont
  {Tagawa}}, \bibinfo {author} {\bibfnamefont {Z.}~\bibnamefont {Haiman}}, \
  and\ \bibinfo {author} {\bibfnamefont {B.}~\bibnamefont {Kocsis}},\ }\href
  {\doibase 10.3847/1538-4357/ab9b8c} {\bibfield  {journal} {\bibinfo
  {journal} {Astrophys. J.}\ }\textbf {\bibinfo {volume} {898}},\ \bibinfo
  {pages} {25} (\bibinfo {year} {2020})},\ \Eprint
  {http://arxiv.org/abs/1912.08218} {arXiv:1912.08218 [astro-ph.GA]}
  \BibitemShut {NoStop}%
\bibitem [{\citenamefont {Zevin}\ \emph {et~al.}(2019)\citenamefont {Zevin},
  \citenamefont {Samsing}, \citenamefont {Rodriguez}, \citenamefont {Haster},\
  and\ \citenamefont {Ramirez-Ruiz}}]{Zevin:2018kzq}%
  \BibitemOpen
  \bibfield  {author} {\bibinfo {author} {\bibfnamefont {M.}~\bibnamefont
  {Zevin}}, \bibinfo {author} {\bibfnamefont {J.}~\bibnamefont {Samsing}},
  \bibinfo {author} {\bibfnamefont {C.}~\bibnamefont {Rodriguez}}, \bibinfo
  {author} {\bibfnamefont {C.-J.}\ \bibnamefont {Haster}}, \ and\ \bibinfo
  {author} {\bibfnamefont {E.}~\bibnamefont {Ramirez-Ruiz}},\ }\href {\doibase
  10.3847/1538-4357/aaf6ec} {\bibfield  {journal} {\bibinfo  {journal}
  {Astrophys. J.}\ }\textbf {\bibinfo {volume} {871}},\ \bibinfo {pages} {91}
  (\bibinfo {year} {2019})},\ \Eprint {http://arxiv.org/abs/1810.00901}
  {arXiv:1810.00901 [astro-ph.HE]} \BibitemShut {NoStop}%
\bibitem [{\citenamefont {Samsing}\ and\ \citenamefont
  {D'Orazio}(2018)}]{Samsing:2018isx}%
  \BibitemOpen
  \bibfield  {author} {\bibinfo {author} {\bibfnamefont {J.}~\bibnamefont
  {Samsing}}\ and\ \bibinfo {author} {\bibfnamefont {D.~J.}\ \bibnamefont
  {D'Orazio}},\ }\href {\doibase 10.1093/mnras/sty2334} {\bibfield  {journal}
  {\bibinfo  {journal} {Mon. Not. Roy. Astron. Soc.}\ }\textbf {\bibinfo
  {volume} {481}},\ \bibinfo {pages} {5445} (\bibinfo {year} {2018})},\ \Eprint
  {http://arxiv.org/abs/1804.06519} {arXiv:1804.06519 [astro-ph.HE]}
  \BibitemShut {NoStop}%
\bibitem [{\citenamefont {O'Leary}\ \emph {et~al.}(2009)\citenamefont
  {O'Leary}, \citenamefont {Kocsis},\ and\ \citenamefont
  {Loeb}}]{10.1111/j.1365-2966.2009.14653.x}%
  \BibitemOpen
  \bibfield  {author} {\bibinfo {author} {\bibfnamefont {R.~M.}\ \bibnamefont
  {O'Leary}}, \bibinfo {author} {\bibfnamefont {B.}~\bibnamefont {Kocsis}}, \
  and\ \bibinfo {author} {\bibfnamefont {A.}~\bibnamefont {Loeb}},\ }\href
  {\doibase 10.1111/j.1365-2966.2009.14653.x} {\bibfield  {journal} {\bibinfo
  {journal} {Monthly Notices of the Royal Astronomical Society}\ }\textbf
  {\bibinfo {volume} {395}},\ \bibinfo {pages} {2127} (\bibinfo {year}
  {2009})},\ \Eprint
  {http://arxiv.org/abs/https://academic.oup.com/mnras/article-pdf/395/4/2127/2931749/mnras0395-2127.pdf}
  {https://academic.oup.com/mnras/article-pdf/395/4/2127/2931749/mnras0395-2127.pdf}
  \BibitemShut {NoStop}%
\bibitem [{\citenamefont {Amaro-Seoane}(2018)}]{Amaro-Seoane:2018gbb}%
  \BibitemOpen
  \bibfield  {author} {\bibinfo {author} {\bibfnamefont {P.}~\bibnamefont
  {Amaro-Seoane}},\ }\href {\doibase 10.1103/PhysRevD.98.063018} {\bibfield
  {journal} {\bibinfo  {journal} {Phys. Rev. D}\ }\textbf {\bibinfo {volume}
  {98}},\ \bibinfo {pages} {063018} (\bibinfo {year} {2018})},\ \Eprint
  {http://arxiv.org/abs/1807.03824} {arXiv:1807.03824 [astro-ph.HE]}
  \BibitemShut {NoStop}%
\bibitem [{\citenamefont {East}\ \emph {et~al.}(2013)\citenamefont {East},
  \citenamefont {McWilliams}, \citenamefont {Levin},\ and\ \citenamefont
  {Pretorius}}]{East:2012xq}%
  \BibitemOpen
  \bibfield  {author} {\bibinfo {author} {\bibfnamefont {W.~E.}\ \bibnamefont
  {East}}, \bibinfo {author} {\bibfnamefont {S.~T.}\ \bibnamefont
  {McWilliams}}, \bibinfo {author} {\bibfnamefont {J.}~\bibnamefont {Levin}}, \
  and\ \bibinfo {author} {\bibfnamefont {F.}~\bibnamefont {Pretorius}},\ }\href
  {\doibase 10.1103/PhysRevD.87.043004} {\bibfield  {journal} {\bibinfo
  {journal} {Phys. Rev.}\ }\textbf {\bibinfo {volume} {D87}},\ \bibinfo {pages}
  {043004} (\bibinfo {year} {2013})},\ \Eprint {http://arxiv.org/abs/1212.0837}
  {arXiv:1212.0837 [gr-qc]} \BibitemShut {NoStop}%
\bibitem [{\citenamefont {Loutrel}(2020)}]{Loutrel:2020kmm}%
  \BibitemOpen
  \bibfield  {author} {\bibinfo {author} {\bibfnamefont {N.}~\bibnamefont
  {Loutrel}},\ }\href@noop {} {\  (\bibinfo {year} {2020})},\ \Eprint
  {http://arxiv.org/abs/2009.11332} {arXiv:2009.11332 [gr-qc]} \BibitemShut
  {NoStop}%
\bibitem [{\citenamefont {Huerta}\ \emph {et~al.}(2018)\citenamefont {Huerta}
  \emph {et~al.}}]{Huerta:2017kez}%
  \BibitemOpen
  \bibfield  {author} {\bibinfo {author} {\bibfnamefont {E.~A.}\ \bibnamefont
  {Huerta}} \emph {et~al.},\ }\href {\doibase 10.1103/PhysRevD.97.024031}
  {\bibfield  {journal} {\bibinfo  {journal} {Phys. Rev.}\ }\textbf {\bibinfo
  {volume} {D97}},\ \bibinfo {pages} {024031} (\bibinfo {year} {2018})},\
  \Eprint {http://arxiv.org/abs/1711.06276} {arXiv:1711.06276 [gr-qc]}
  \BibitemShut {NoStop}%
\bibitem [{\citenamefont {Cao}\ and\ \citenamefont {Han}(2017)}]{Cao:2017ndf}%
  \BibitemOpen
  \bibfield  {author} {\bibinfo {author} {\bibfnamefont {Z.}~\bibnamefont
  {Cao}}\ and\ \bibinfo {author} {\bibfnamefont {W.-B.}\ \bibnamefont {Han}},\
  }\href {\doibase 10.1103/PhysRevD.96.044028} {\bibfield  {journal} {\bibinfo
  {journal} {Phys. Rev.}\ }\textbf {\bibinfo {volume} {D96}},\ \bibinfo {pages}
  {044028} (\bibinfo {year} {2017})},\ \Eprint
  {http://arxiv.org/abs/1708.00166} {arXiv:1708.00166 [gr-qc]} \BibitemShut
  {NoStop}%
\bibitem [{\citenamefont {Liu}\ \emph {et~al.}(2019)\citenamefont {Liu},
  \citenamefont {Cao},\ and\ \citenamefont {Shao}}]{Liu:2019jpg}%
  \BibitemOpen
  \bibfield  {author} {\bibinfo {author} {\bibfnamefont {X.}~\bibnamefont
  {Liu}}, \bibinfo {author} {\bibfnamefont {Z.}~\bibnamefont {Cao}}, \ and\
  \bibinfo {author} {\bibfnamefont {L.}~\bibnamefont {Shao}},\ }\href@noop {}
  {\  (\bibinfo {year} {2019})},\ \Eprint {http://arxiv.org/abs/1910.00784}
  {arXiv:1910.00784 [gr-qc]} \BibitemShut {NoStop}%
\bibitem [{\citenamefont {Gold}\ and\ \citenamefont
  {Br{\"u}gmann}(2013)}]{Gold:2012tk}%
  \BibitemOpen
  \bibfield  {author} {\bibinfo {author} {\bibfnamefont {R.}~\bibnamefont
  {Gold}}\ and\ \bibinfo {author} {\bibfnamefont {B.}~\bibnamefont
  {Br{\"u}gmann}},\ }\href {\doibase 10.1103/PhysRevD.88.064051} {\bibfield
  {journal} {\bibinfo  {journal} {Phys. Rev.}\ }\textbf {\bibinfo {volume}
  {D88}},\ \bibinfo {pages} {064051} (\bibinfo {year} {2013})},\ \Eprint
  {http://arxiv.org/abs/1209.4085} {arXiv:1209.4085 [gr-qc]} \BibitemShut
  {NoStop}%
\bibitem [{\citenamefont {Nelson}\ \emph {et~al.}(2019)\citenamefont {Nelson},
  \citenamefont {Etienne}, \citenamefont {McWilliams},\ and\ \citenamefont
  {Nguyen}}]{Nelson:2019czq}%
  \BibitemOpen
  \bibfield  {author} {\bibinfo {author} {\bibfnamefont {P.~E.}\ \bibnamefont
  {Nelson}}, \bibinfo {author} {\bibfnamefont {Z.~B.}\ \bibnamefont {Etienne}},
  \bibinfo {author} {\bibfnamefont {S.~T.}\ \bibnamefont {McWilliams}}, \ and\
  \bibinfo {author} {\bibfnamefont {V.}~\bibnamefont {Nguyen}},\ }\href
  {\doibase 10.1103/PhysRevD.100.124045} {\bibfield  {journal} {\bibinfo
  {journal} {Phys. Rev. D}\ }\textbf {\bibinfo {volume} {100}},\ \bibinfo
  {pages} {124045} (\bibinfo {year} {2019})},\ \Eprint
  {http://arxiv.org/abs/1909.08621} {arXiv:1909.08621 [gr-qc]} \BibitemShut
  {NoStop}%
\bibitem [{\citenamefont {Bae}\ \emph {et~al.}(2020)\citenamefont {Bae},
  \citenamefont {Lee},\ and\ \citenamefont {Kang}}]{Bae:2020hla}%
  \BibitemOpen
  \bibfield  {author} {\bibinfo {author} {\bibfnamefont {Y.-B.}\ \bibnamefont
  {Bae}}, \bibinfo {author} {\bibfnamefont {H.~M.}\ \bibnamefont {Lee}}, \ and\
  \bibinfo {author} {\bibfnamefont {G.}~\bibnamefont {Kang}},\ }\href@noop {}
  {\  (\bibinfo {year} {2020})},\ \Eprint {http://arxiv.org/abs/2007.14019}
  {arXiv:2007.14019 [gr-qc]} \BibitemShut {NoStop}%
\bibitem [{\citenamefont {Damour}\ \emph {et~al.}(2014)\citenamefont {Damour},
  \citenamefont {Guercilena}, \citenamefont {Hinder}, \citenamefont {Hopper},
  \citenamefont {Nagar} \emph {et~al.}}]{Damour:2014afa}%
  \BibitemOpen
  \bibfield  {author} {\bibinfo {author} {\bibfnamefont {T.}~\bibnamefont
  {Damour}}, \bibinfo {author} {\bibfnamefont {F.}~\bibnamefont {Guercilena}},
  \bibinfo {author} {\bibfnamefont {I.}~\bibnamefont {Hinder}}, \bibinfo
  {author} {\bibfnamefont {S.}~\bibnamefont {Hopper}}, \bibinfo {author}
  {\bibfnamefont {A.}~\bibnamefont {Nagar}},  \emph {et~al.},\ }\href@noop {}
  {\  (\bibinfo {year} {2014})},\ \Eprint {http://arxiv.org/abs/1402.7307}
  {arXiv:1402.7307 [gr-qc]} \BibitemShut {NoStop}%
\bibitem [{\citenamefont {Buonanno}\ and\ \citenamefont
  {Damour}(1999)}]{Buonanno:1998gg}%
  \BibitemOpen
  \bibfield  {author} {\bibinfo {author} {\bibfnamefont {A.}~\bibnamefont
  {Buonanno}}\ and\ \bibinfo {author} {\bibfnamefont {T.}~\bibnamefont
  {Damour}},\ }\href {\doibase 10.1103/PhysRevD.59.084006} {\bibfield
  {journal} {\bibinfo  {journal} {Phys. Rev.}\ }\textbf {\bibinfo {volume}
  {D59}},\ \bibinfo {pages} {084006} (\bibinfo {year} {1999})},\ \Eprint
  {http://arxiv.org/abs/gr-qc/9811091} {arXiv:gr-qc/9811091} \BibitemShut
  {NoStop}%
\bibitem [{\citenamefont {Buonanno}\ and\ \citenamefont
  {Damour}(2000)}]{Buonanno:2000ef}%
  \BibitemOpen
  \bibfield  {author} {\bibinfo {author} {\bibfnamefont {A.}~\bibnamefont
  {Buonanno}}\ and\ \bibinfo {author} {\bibfnamefont {T.}~\bibnamefont
  {Damour}},\ }\href {\doibase 10.1103/PhysRevD.62.064015} {\bibfield
  {journal} {\bibinfo  {journal} {Phys. Rev.}\ }\textbf {\bibinfo {volume}
  {D62}},\ \bibinfo {pages} {064015} (\bibinfo {year} {2000})},\ \Eprint
  {http://arxiv.org/abs/gr-qc/0001013} {arXiv:gr-qc/0001013} \BibitemShut
  {NoStop}%
\bibitem [{\citenamefont {Damour}\ \emph {et~al.}(2000)\citenamefont {Damour},
  \citenamefont {Jaranowski},\ and\ \citenamefont {Schaefer}}]{Damour:2000we}%
  \BibitemOpen
  \bibfield  {author} {\bibinfo {author} {\bibfnamefont {T.}~\bibnamefont
  {Damour}}, \bibinfo {author} {\bibfnamefont {P.}~\bibnamefont {Jaranowski}},
  \ and\ \bibinfo {author} {\bibfnamefont {G.}~\bibnamefont {Schaefer}},\
  }\href {\doibase 10.1103/PhysRevD.62.084011} {\bibfield  {journal} {\bibinfo
  {journal} {Phys. Rev.}\ }\textbf {\bibinfo {volume} {D62}},\ \bibinfo {pages}
  {084011} (\bibinfo {year} {2000})},\ \Eprint
  {http://arxiv.org/abs/gr-qc/0005034} {arXiv:gr-qc/0005034 [gr-qc]}
  \BibitemShut {NoStop}%
\bibitem [{\citenamefont {Damour}(2001)}]{Damour:2001tu}%
  \BibitemOpen
  \bibfield  {author} {\bibinfo {author} {\bibfnamefont {T.}~\bibnamefont
  {Damour}},\ }\href {\doibase 10.1103/PhysRevD.64.124013} {\bibfield
  {journal} {\bibinfo  {journal} {Phys. Rev.}\ }\textbf {\bibinfo {volume}
  {D64}},\ \bibinfo {pages} {124013} (\bibinfo {year} {2001})},\ \Eprint
  {http://arxiv.org/abs/gr-qc/0103018} {arXiv:gr-qc/0103018} \BibitemShut
  {NoStop}%
\bibitem [{\citenamefont {Damour}\ \emph {et~al.}(2008)\citenamefont {Damour},
  \citenamefont {Jaranowski},\ and\ \citenamefont
  {Sch{\"a}fer}}]{Damour:2008qf}%
  \BibitemOpen
  \bibfield  {author} {\bibinfo {author} {\bibfnamefont {T.}~\bibnamefont
  {Damour}}, \bibinfo {author} {\bibfnamefont {P.}~\bibnamefont {Jaranowski}},
  \ and\ \bibinfo {author} {\bibfnamefont {G.}~\bibnamefont {Sch{\"a}fer}},\
  }\href {\doibase 10.1103/PhysRevD.78.024009} {\bibfield  {journal} {\bibinfo
  {journal} {Phys.Rev.}\ }\textbf {\bibinfo {volume} {D78}},\ \bibinfo {pages}
  {024009} (\bibinfo {year} {2008})},\ \Eprint {http://arxiv.org/abs/0803.0915}
  {arXiv:0803.0915 [gr-qc]} \BibitemShut {NoStop}%
\bibitem [{\citenamefont {Nagar}(2011)}]{Nagar:2011fx}%
  \BibitemOpen
  \bibfield  {author} {\bibinfo {author} {\bibfnamefont {A.}~\bibnamefont
  {Nagar}},\ }\href {\doibase 10.1103/PhysRevD.84.084028} {\bibfield  {journal}
  {\bibinfo  {journal} {Phys.Rev.}\ }\textbf {\bibinfo {volume} {D84}},\
  \bibinfo {pages} {084028} (\bibinfo {year} {2011})},\ \Eprint
  {http://arxiv.org/abs/1106.4349} {arXiv:1106.4349 [gr-qc]} \BibitemShut
  {NoStop}%
\bibitem [{\citenamefont {Damour}\ \emph {et~al.}(2015)\citenamefont {Damour},
  \citenamefont {Jaranowski},\ and\ \citenamefont {Schäfer}}]{Damour:2015isa}%
  \BibitemOpen
  \bibfield  {author} {\bibinfo {author} {\bibfnamefont {T.}~\bibnamefont
  {Damour}}, \bibinfo {author} {\bibfnamefont {P.}~\bibnamefont {Jaranowski}},
  \ and\ \bibinfo {author} {\bibfnamefont {G.}~\bibnamefont {Schäfer}},\
  }\href {\doibase 10.1103/PhysRevD.91.084024} {\bibfield  {journal} {\bibinfo
  {journal} {Phys. Rev.}\ }\textbf {\bibinfo {volume} {D91}},\ \bibinfo {pages}
  {084024} (\bibinfo {year} {2015})},\ \Eprint
  {http://arxiv.org/abs/1502.07245} {arXiv:1502.07245 [gr-qc]} \BibitemShut
  {NoStop}%
\bibitem [{teo()}]{teobresums}%
  \BibitemOpen
  \href@noop {} {}\bibinfo {howpublished}
  {\url{https://bitbucket.org/eob_ihes/teobresums/src/master/}},\ \bibinfo
  {note} {{TEOBResumS code.}}\BibitemShut {Stop}%
\bibitem [{\citenamefont {Chiaramello}\ and\ \citenamefont
  {Nagar}(2020)}]{Chiaramello:2020ehz}%
  \BibitemOpen
  \bibfield  {author} {\bibinfo {author} {\bibfnamefont {D.}~\bibnamefont
  {Chiaramello}}\ and\ \bibinfo {author} {\bibfnamefont {A.}~\bibnamefont
  {Nagar}},\ }\href {\doibase 10.1103/PhysRevD.101.101501} {\bibfield
  {journal} {\bibinfo  {journal} {Phys. Rev. D}\ }\textbf {\bibinfo {volume}
  {101}},\ \bibinfo {pages} {101501} (\bibinfo {year} {2020})},\ \Eprint
  {http://arxiv.org/abs/2001.11736} {arXiv:2001.11736 [gr-qc]} \BibitemShut
  {NoStop}%
\bibitem [{\citenamefont {Nagar}\ \emph {et~al.}(2017)\citenamefont {Nagar},
  \citenamefont {Riemenschneider},\ and\ \citenamefont
  {Pratten}}]{Nagar:2017jdw}%
  \BibitemOpen
  \bibfield  {author} {\bibinfo {author} {\bibfnamefont {A.}~\bibnamefont
  {Nagar}}, \bibinfo {author} {\bibfnamefont {G.}~\bibnamefont
  {Riemenschneider}}, \ and\ \bibinfo {author} {\bibfnamefont {G.}~\bibnamefont
  {Pratten}},\ }\href {\doibase 10.1103/PhysRevD.96.084045} {\bibfield
  {journal} {\bibinfo  {journal} {Phys. Rev.}\ }\textbf {\bibinfo {volume}
  {D96}},\ \bibinfo {pages} {084045} (\bibinfo {year} {2017})},\ \Eprint
  {http://arxiv.org/abs/1703.06814} {arXiv:1703.06814 [gr-qc]} \BibitemShut
  {NoStop}%
\bibitem [{\citenamefont {Nagar}\ \emph {et~al.}(2019)\citenamefont {Nagar},
  \citenamefont {Pratten}, \citenamefont {Riemenschneider},\ and\ \citenamefont
  {Gamba}}]{Nagar:2019wds}%
  \BibitemOpen
  \bibfield  {author} {\bibinfo {author} {\bibfnamefont {A.}~\bibnamefont
  {Nagar}}, \bibinfo {author} {\bibfnamefont {G.}~\bibnamefont {Pratten}},
  \bibinfo {author} {\bibfnamefont {G.}~\bibnamefont {Riemenschneider}}, \ and\
  \bibinfo {author} {\bibfnamefont {R.}~\bibnamefont {Gamba}},\ }\href@noop {}
  {\  (\bibinfo {year} {2019})},\ \Eprint {http://arxiv.org/abs/1904.09550}
  {arXiv:1904.09550 [gr-qc]} \BibitemShut {NoStop}%
\bibitem [{\citenamefont {Nagar}\ \emph {et~al.}(2020)\citenamefont {Nagar},
  \citenamefont {Riemenschneider}, \citenamefont {Pratten}, \citenamefont
  {Rettegno},\ and\ \citenamefont {Messina}}]{Nagar:2020pcj}%
  \BibitemOpen
  \bibfield  {author} {\bibinfo {author} {\bibfnamefont {A.}~\bibnamefont
  {Nagar}}, \bibinfo {author} {\bibfnamefont {G.}~\bibnamefont
  {Riemenschneider}}, \bibinfo {author} {\bibfnamefont {G.}~\bibnamefont
  {Pratten}}, \bibinfo {author} {\bibfnamefont {P.}~\bibnamefont {Rettegno}}, \
  and\ \bibinfo {author} {\bibfnamefont {F.}~\bibnamefont {Messina}},\
  }\href@noop {} {\  (\bibinfo {year} {2020})},\ \Eprint
  {http://arxiv.org/abs/2001.09082} {arXiv:2001.09082 [gr-qc]} \BibitemShut
  {NoStop}%
\bibitem [{\citenamefont {Bini}\ \emph {et~al.}(2019)\citenamefont {Bini},
  \citenamefont {Damour},\ and\ \citenamefont {Geralico}}]{Bini:2019nra}%
  \BibitemOpen
  \bibfield  {author} {\bibinfo {author} {\bibfnamefont {D.}~\bibnamefont
  {Bini}}, \bibinfo {author} {\bibfnamefont {T.}~\bibnamefont {Damour}}, \ and\
  \bibinfo {author} {\bibfnamefont {A.}~\bibnamefont {Geralico}},\ }\href
  {\doibase 10.1103/PhysRevLett.123.231104} {\bibfield  {journal} {\bibinfo
  {journal} {Phys. Rev. Lett.}\ }\textbf {\bibinfo {volume} {123}},\ \bibinfo
  {pages} {231104} (\bibinfo {year} {2019})},\ \Eprint
  {http://arxiv.org/abs/1909.02375} {arXiv:1909.02375 [gr-qc]} \BibitemShut
  {NoStop}%
\bibitem [{\citenamefont {Bini}\ \emph
  {et~al.}(2020{\natexlab{a}})\citenamefont {Bini}, \citenamefont {Damour},\
  and\ \citenamefont {Geralico}}]{Bini:2020wpo}%
  \BibitemOpen
  \bibfield  {author} {\bibinfo {author} {\bibfnamefont {D.}~\bibnamefont
  {Bini}}, \bibinfo {author} {\bibfnamefont {T.}~\bibnamefont {Damour}}, \ and\
  \bibinfo {author} {\bibfnamefont {A.}~\bibnamefont {Geralico}},\ }\href
  {\doibase 10.1103/PhysRevD.102.024062} {\bibfield  {journal} {\bibinfo
  {journal} {Phys. Rev. D}\ }\textbf {\bibinfo {volume} {102}},\ \bibinfo
  {pages} {024062} (\bibinfo {year} {2020}{\natexlab{a}})},\ \Eprint
  {http://arxiv.org/abs/2003.11891} {arXiv:2003.11891 [gr-qc]} \BibitemShut
  {NoStop}%
\bibitem [{\citenamefont {Bini}\ \emph
  {et~al.}(2020{\natexlab{b}})\citenamefont {Bini}, \citenamefont {Damour},\
  and\ \citenamefont {Geralico}}]{Bini:2020nsb}%
  \BibitemOpen
  \bibfield  {author} {\bibinfo {author} {\bibfnamefont {D.}~\bibnamefont
  {Bini}}, \bibinfo {author} {\bibfnamefont {T.}~\bibnamefont {Damour}}, \ and\
  \bibinfo {author} {\bibfnamefont {A.}~\bibnamefont {Geralico}},\ }\href
  {\doibase 10.1103/PhysRevD.102.024061} {\bibfield  {journal} {\bibinfo
  {journal} {Phys. Rev. D}\ }\textbf {\bibinfo {volume} {102}},\ \bibinfo
  {pages} {024061} (\bibinfo {year} {2020}{\natexlab{b}})},\ \Eprint
  {http://arxiv.org/abs/2004.05407} {arXiv:2004.05407 [gr-qc]} \BibitemShut
  {NoStop}%
\bibitem [{\citenamefont {Bini}\ \emph
  {et~al.}(2020{\natexlab{c}})\citenamefont {Bini}, \citenamefont {Damour},\
  and\ \citenamefont {Geralico}}]{Bini:2020hmy}%
  \BibitemOpen
  \bibfield  {author} {\bibinfo {author} {\bibfnamefont {D.}~\bibnamefont
  {Bini}}, \bibinfo {author} {\bibfnamefont {T.}~\bibnamefont {Damour}}, \ and\
  \bibinfo {author} {\bibfnamefont {A.}~\bibnamefont {Geralico}},\ }\href@noop
  {} {\  (\bibinfo {year} {2020}{\natexlab{c}})},\ \Eprint
  {http://arxiv.org/abs/2007.11239} {arXiv:2007.11239 [gr-qc]} \BibitemShut
  {NoStop}%
\bibitem [{\citenamefont {Nagar}\ \emph {et~al.}(2018)\citenamefont {Nagar}
  \emph {et~al.}}]{Nagar:2018zoe}%
  \BibitemOpen
  \bibfield  {author} {\bibinfo {author} {\bibfnamefont {A.}~\bibnamefont
  {Nagar}} \emph {et~al.},\ }\href {\doibase 10.1103/PhysRevD.98.104052}
  {\bibfield  {journal} {\bibinfo  {journal} {Phys. Rev.}\ }\textbf {\bibinfo
  {volume} {D98}},\ \bibinfo {pages} {104052} (\bibinfo {year} {2018})},\
  \Eprint {http://arxiv.org/abs/1806.01772} {arXiv:1806.01772 [gr-qc]}
  \BibitemShut {NoStop}%
\bibitem [{\citenamefont {Damour}\ and\ \citenamefont
  {Nagar}(2014{\natexlab{a}})}]{Damour:2014yha}%
  \BibitemOpen
  \bibfield  {author} {\bibinfo {author} {\bibfnamefont {T.}~\bibnamefont
  {Damour}}\ and\ \bibinfo {author} {\bibfnamefont {A.}~\bibnamefont {Nagar}},\
  }\href {\doibase 10.1103/PhysRevD.90.024054} {\bibfield  {journal} {\bibinfo
  {journal} {Phys.Rev.}\ }\textbf {\bibinfo {volume} {D90}},\ \bibinfo {pages}
  {024054} (\bibinfo {year} {2014}{\natexlab{a}})},\ \Eprint
  {http://arxiv.org/abs/1406.0401} {arXiv:1406.0401 [gr-qc]} \BibitemShut
  {NoStop}%
\bibitem [{\citenamefont {Damour}\ and\ \citenamefont
  {Nagar}(2014{\natexlab{b}})}]{Damour:2014sva}%
  \BibitemOpen
  \bibfield  {author} {\bibinfo {author} {\bibfnamefont {T.}~\bibnamefont
  {Damour}}\ and\ \bibinfo {author} {\bibfnamefont {A.}~\bibnamefont {Nagar}},\
  }\href {\doibase 10.1103/PhysRevD.90.044018} {\bibfield  {journal} {\bibinfo
  {journal} {Phys.Rev.}\ }\textbf {\bibinfo {volume} {D90}},\ \bibinfo {pages}
  {044018} (\bibinfo {year} {2014}{\natexlab{b}})},\ \Eprint
  {http://arxiv.org/abs/1406.6913} {arXiv:1406.6913 [gr-qc]} \BibitemShut
  {NoStop}%
\bibitem [{\citenamefont {Kidder}(2008)}]{Kidder:2007rt}%
  \BibitemOpen
  \bibfield  {author} {\bibinfo {author} {\bibfnamefont {L.~E.}\ \bibnamefont
  {Kidder}},\ }\href {\doibase 10.1103/PhysRevD.77.044016} {\bibfield
  {journal} {\bibinfo  {journal} {Phys. Rev.}\ }\textbf {\bibinfo {volume}
  {D77}},\ \bibinfo {pages} {044016} (\bibinfo {year} {2008})},\ \Eprint
  {http://arxiv.org/abs/0710.0614} {arXiv:0710.0614 [gr-qc]} \BibitemShut
  {NoStop}%
\bibitem [{\citenamefont {Harms}\ \emph {et~al.}(2014)\citenamefont {Harms},
  \citenamefont {Bernuzzi}, \citenamefont {Nagar},\ and\ \citenamefont
  {Zenginoglu}}]{Harms:2014dqa}%
  \BibitemOpen
  \bibfield  {author} {\bibinfo {author} {\bibfnamefont {E.}~\bibnamefont
  {Harms}}, \bibinfo {author} {\bibfnamefont {S.}~\bibnamefont {Bernuzzi}},
  \bibinfo {author} {\bibfnamefont {A.}~\bibnamefont {Nagar}}, \ and\ \bibinfo
  {author} {\bibfnamefont {A.}~\bibnamefont {Zenginoglu}},\ }\href {\doibase
  10.1088/0264-9381/31/24/245004} {\bibfield  {journal} {\bibinfo  {journal}
  {Class.Quant.Grav.}\ }\textbf {\bibinfo {volume} {31}},\ \bibinfo {pages}
  {245004} (\bibinfo {year} {2014})},\ \Eprint {http://arxiv.org/abs/1406.5983}
  {arXiv:1406.5983 [gr-qc]} \BibitemShut {NoStop}%
\bibitem [{\citenamefont {Damour}\ \emph {et~al.}(2013)\citenamefont {Damour},
  \citenamefont {Nagar},\ and\ \citenamefont {Bernuzzi}}]{Damour:2012ky}%
  \BibitemOpen
  \bibfield  {author} {\bibinfo {author} {\bibfnamefont {T.}~\bibnamefont
  {Damour}}, \bibinfo {author} {\bibfnamefont {A.}~\bibnamefont {Nagar}}, \
  and\ \bibinfo {author} {\bibfnamefont {S.}~\bibnamefont {Bernuzzi}},\ }\href
  {\doibase 10.1103/PhysRevD.87.084035} {\bibfield  {journal} {\bibinfo
  {journal} {Phys.Rev.}\ }\textbf {\bibinfo {volume} {D87}},\ \bibinfo {pages}
  {084035} (\bibinfo {year} {2013})},\ \Eprint {http://arxiv.org/abs/1212.4357}
  {arXiv:1212.4357 [gr-qc]} \BibitemShut {NoStop}%
\bibitem [{\citenamefont {Bini}\ and\ \citenamefont
  {Damour}(2013)}]{Bini:2013zaa}%
  \BibitemOpen
  \bibfield  {author} {\bibinfo {author} {\bibfnamefont {D.}~\bibnamefont
  {Bini}}\ and\ \bibinfo {author} {\bibfnamefont {T.}~\bibnamefont {Damour}},\
  }\href {\doibase 10.1103/PhysRevD.87.121501} {\bibfield  {journal} {\bibinfo
  {journal} {Phys.Rev.}\ }\textbf {\bibinfo {volume} {D87}},\ \bibinfo {pages}
  {121501} (\bibinfo {year} {2013})},\ \Eprint {http://arxiv.org/abs/1305.4884}
  {arXiv:1305.4884 [gr-qc]} \BibitemShut {NoStop}%
\bibitem [{\citenamefont {Damour}\ \emph {et~al.}(2016)\citenamefont {Damour},
  \citenamefont {Jaranowski},\ and\ \citenamefont
  {Sch{\"a}fer}}]{Damour:2016abl}%
  \BibitemOpen
  \bibfield  {author} {\bibinfo {author} {\bibfnamefont {T.}~\bibnamefont
  {Damour}}, \bibinfo {author} {\bibfnamefont {P.}~\bibnamefont {Jaranowski}},
  \ and\ \bibinfo {author} {\bibfnamefont {G.}~\bibnamefont {Sch{\"a}fer}},\
  }\href {\doibase 10.1103/PhysRevD.93.084014} {\bibfield  {journal} {\bibinfo
  {journal} {Phys. Rev.}\ }\textbf {\bibinfo {volume} {D93}},\ \bibinfo {pages}
  {084014} (\bibinfo {year} {2016})},\ \Eprint
  {http://arxiv.org/abs/1601.01283} {arXiv:1601.01283 [gr-qc]} \BibitemShut
  {NoStop}%
\bibitem [{\citenamefont {Abbott}\ \emph
  {et~al.}(2020{\natexlab{a}})\citenamefont {Abbott} \emph
  {et~al.}}]{Abbott:2020tfl}%
  \BibitemOpen
  \bibfield  {author} {\bibinfo {author} {\bibfnamefont {R.}~\bibnamefont
  {Abbott}} \emph {et~al.} (\bibinfo {collaboration} {LIGO Scientific,
  Virgo}),\ }\href {\doibase 10.1103/PhysRevLett.125.101102} {\bibfield
  {journal} {\bibinfo  {journal} {Phys. Rev. Lett.}\ }\textbf {\bibinfo
  {volume} {125}},\ \bibinfo {pages} {101102} (\bibinfo {year}
  {2020}{\natexlab{a}})},\ \Eprint {http://arxiv.org/abs/2009.01075}
  {arXiv:2009.01075 [gr-qc]} \BibitemShut {NoStop}%
\bibitem [{\citenamefont {Abbott}\ \emph
  {et~al.}(2020{\natexlab{b}})\citenamefont {Abbott} \emph
  {et~al.}}]{Abbott:2020mjq}%
  \BibitemOpen
  \bibfield  {author} {\bibinfo {author} {\bibfnamefont {R.}~\bibnamefont
  {Abbott}} \emph {et~al.} (\bibinfo {collaboration} {LIGO Scientific,
  Virgo}),\ }\href {\doibase 10.3847/2041-8213/aba493} {\bibfield  {journal}
  {\bibinfo  {journal} {Astrophys. J. Lett.}\ }\textbf {\bibinfo {volume}
  {900}},\ \bibinfo {pages} {L13} (\bibinfo {year} {2020}{\natexlab{b}})},\
  \Eprint {http://arxiv.org/abs/2009.01190} {arXiv:2009.01190 [astro-ph.HE]}
  \BibitemShut {NoStop}%
\bibitem [{\citenamefont {Calder\F3n~Bustillo}\ \emph
  {et~al.}(2020)\citenamefont {Calder\F3n~Bustillo}, \citenamefont
  {Sanchis-Gual}, \citenamefont {Torres-Forn\E9},\ and\ \citenamefont
  {Font}}]{CalderonBustillo:2020odh}%
  \BibitemOpen
  \bibfield  {author} {\bibinfo {author} {\bibfnamefont {J.}~\bibnamefont
  {Calder\F3n~Bustillo}}, \bibinfo {author} {\bibfnamefont {N.}~\bibnamefont
  {Sanchis-Gual}}, \bibinfo {author} {\bibfnamefont {A.}~\bibnamefont
  {Torres-Forn\E9}}, \ and\ \bibinfo {author} {\bibfnamefont {J.~A.}\
  \bibnamefont {Font}},\ }\href@noop {} {\  (\bibinfo {year} {2020})},\ \Eprint
  {http://arxiv.org/abs/2009.01066} {arXiv:2009.01066 [gr-qc]} \BibitemShut
  {NoStop}%
\bibitem [{\citenamefont {Romero-Shaw}\ \emph {et~al.}(2020)\citenamefont
  {Romero-Shaw}, \citenamefont {Lasky}, \citenamefont {Thrane},\ and\
  \citenamefont {Bustillo}}]{Romero-Shaw:2020thy}%
  \BibitemOpen
  \bibfield  {author} {\bibinfo {author} {\bibfnamefont {I.~M.}\ \bibnamefont
  {Romero-Shaw}}, \bibinfo {author} {\bibfnamefont {P.~D.}\ \bibnamefont
  {Lasky}}, \bibinfo {author} {\bibfnamefont {E.}~\bibnamefont {Thrane}}, \
  and\ \bibinfo {author} {\bibfnamefont {J.~C.}\ \bibnamefont {Bustillo}},\
  }\href@noop {} {\  (\bibinfo {year} {2020})},\ \Eprint
  {http://arxiv.org/abs/2009.04771} {arXiv:2009.04771 [astro-ph.HE]}
  \BibitemShut {NoStop}%
\bibitem [{\citenamefont {Gayathri}\ \emph {et~al.}(2020)\citenamefont
  {Gayathri}, \citenamefont {Healy}, \citenamefont {Lange}, \citenamefont
  {O'Brien}, \citenamefont {Szczepanczyk}, \citenamefont {Bartos},
  \citenamefont {Campanelli}, \citenamefont {Klimenko}, \citenamefont
  {Lousto},\ and\ \citenamefont {O'Shaughnessy}}]{Gayathri:2020coq}%
  \BibitemOpen
  \bibfield  {author} {\bibinfo {author} {\bibfnamefont {V.}~\bibnamefont
  {Gayathri}}, \bibinfo {author} {\bibfnamefont {J.}~\bibnamefont {Healy}},
  \bibinfo {author} {\bibfnamefont {J.}~\bibnamefont {Lange}}, \bibinfo
  {author} {\bibfnamefont {B.}~\bibnamefont {O'Brien}}, \bibinfo {author}
  {\bibfnamefont {M.}~\bibnamefont {Szczepanczyk}}, \bibinfo {author}
  {\bibfnamefont {I.}~\bibnamefont {Bartos}}, \bibinfo {author} {\bibfnamefont
  {M.}~\bibnamefont {Campanelli}}, \bibinfo {author} {\bibfnamefont
  {S.}~\bibnamefont {Klimenko}}, \bibinfo {author} {\bibfnamefont
  {C.}~\bibnamefont {Lousto}}, \ and\ \bibinfo {author} {\bibfnamefont
  {R.}~\bibnamefont {O'Shaughnessy}},\ }\href@noop {} {\  (\bibinfo {year}
  {2020})},\ \Eprint {http://arxiv.org/abs/2009.05461} {arXiv:2009.05461
  [astro-ph.HE]} \BibitemShut {NoStop}%
\end{thebibliography}%

\end{document}